\documentclass[useAMS,usenatbib,usegraphicx]{mn2e}

\usepackage{ulem}
\usepackage{times}

\title[Search for corannulene in the Red Rectangle]{Search for corannulene (C$_{20}$H$_{10}$) in the Red Rectangle}

\author[P. Pilleri et al.]{P. Pilleri$^{1,}$$^{2}$\thanks{E-mail: paolo.pilleri@cesr.fr}, 
D. Herberth$^3$, 
T. F. Giesen$^3$,
M. Gerin$^4$,
C. Joblin$^{1,}$$^2$,
G. Mulas$^5$,
G. Malloci$^5$,
\newauthor J.-U. Grabow$^6$,
S. Br\"unken$^7$,
L. Surin$^3$,
B. D. Steinberg$^8$,
K. R. Curtis$^8$,
and L. T. Scott$^8$
\\
$^1$Universit\'{e} de Toulouse ;  UPS ;  CESR ;  9 avenue du colonel Roche, F-31028 Toulouse cedex 9, France
\\
$^2$CNRS ; UMR5187 ; F-31028 Toulouse, France
\\
$^3$I. Physikalisches Institut, Universit\"at zu K\"oln, Z\"ulpicher Str. 77, 50937 K\"oln
\\
$^4$LERMA-LRA, CNRS-UMR8112, Observatoire de Paris et \'Ecole Normale Sup\'erieure, 24 rue Lhomond, 75231 Paris Cedex 05, France
\\
$^5$INAF Osservatorio Astronomico di Cagliari, Astrochemistry Group, Strada n.54, Loc. Poggio dei Pini, 09012 Capoterra (CA), Italy
\\
$^6$Gottfried-Willhelm-Leibniz-Universit\"at, Institut f\"ur Physikalische Chemie \& Elektrochemie, Lehrgebiet A, Callinstra{\ss}e 3A, \\ D-30167 Hannover, Germany
\\
$^7$Laboratoire de Chimie Physique Mol\'eculaire (LCPM), \'Ecole Polytechnique F\'ed\'erale de Lausanne (EPFL), Station 6, \\ CH-1015 Lausanne, Switzerland
\\
$^8$Merkert Chemistry Center, Boston College, Chestnut Hill, MA 02467-3860, USA
}

\begin{document}

\date{Accepted 2009 May 6.  Received 2009 May 4; in original form 2009 April 2 }

\pubyear{2009}

\maketitle

\begin{abstract}
Polycyclic Aromatic Hydrocarbons (PAHs) are widely accepted as the carriers of the Aromatic Infrared Bands (AIBs), but an unambiguous identification of any specific interstellar PAH is still missing.
For polar PAHs, pure rotational transitions can be used as spectral fingerprints for identification. Combining dedicated experiments, detailed simulations and observations, we explored the mm wavelength domain to search for specific rotational transitions of corannulene  (C$_{20}$H$_{10}$).
We performed high-resolution spectroscopic measurements and a simulation of the emission spectrum of UV-excited C$_{20}$H$_{10}$ in the environment of the Red Rectangle, calculating its synthetic rotational spectrum. Based on these results, we conducted a first observational campaign at the IRAM 30m telescope towards this source to search for several high-J rotational transitions of C$_{20}$H$_{10}$.
The laboratory detection of the J = 112 $\leftarrow$ 111 transition of corannulene showed that no centrifugal splitting is present up to this line. Observations with the IRAM 30m telescope towards the Red Rectangle do not show any corannulene emission at any of the observed frequencies, down to a rms noise level of T$_{mb} = 8$~mK for the J =135 $\rightarrow$ 134 transition at 137.615 GHz.
Comparing the noise level with the synthetic spectrum, we are able to estimate an upper limit to the fraction of carbon locked in corannulene of about $1.0 \times 10^{-5}$ relative to the total abundance of carbon in PAHs. 
The sensitivity achieved in this work shows that radio spectroscopy can be a powerful tool to search for polar PAHs. We compare this upper limit with models for the PAH size  distribution, emphasising that small PAHs are much less abundant than predicted. We show that this cannot be explained by destruction but is more likely related to the chemistry of their formation in the environment of the Red Rectangle.
\end{abstract}


%

\section{Introduction}	\label{sec:introduction}

Polycyclic Aromatic Hydrocarbons (PAHs) have been proposed more than 20 years ago as an important constituent of the interstellar medium (ISM) \citep{leger84, allamandola85}, being the most likely carriers of the Aromatic Infrared Bands (AIBs), the mid-IR emission features at 3.3, 6.2, 7.7, 8.6, 11.3 and 12.7 $\mu$m that dominate the spectra of many interstellar UV-excited dusty environments \citep{leger89, allamandola89}. PAHs are also thought to be responsible for some of the Diffuse Interstellar Bands (DIBs), more than 300 unidentified absorption features in the UV-visible range observed in the spectra of reddened stars \citep{leger85, vanderzwet85}. Finally, PAHs are nowadays a crucial ingredient in all models of interstellar extinction by dust, playing the role of the "Platt particles" \citep{platt56, donn68} in contributing to the bump at 220~nm and producing the far-UV rise in the extinction curve \citep{li01, cecchipestellini08}.
This has motivated much experimental and theoretical work, but still an unambiguous identification of a single species is missing. 
This task faces the difficulty that bands in the region of the AIBs, which are associated with vibrations of aromatic C-C and C-H bonds, are common to the whole class of PAHs. It is therefore difficult to use these IR bands to identify single species, even though their 
 study led many authors to obtain information on the nature of their carriers \citep[see for example][]{pech02, peeters02,  berne07, joblin08}.  \cite{vijh04, vijh05} proposed  a tentative identification of neutral pyrene (C$_{16}$H$_{10}$) and anthracene (C$_{14}$H$_{10}$) towards the Red Rectangle nebula but this identification was challenged by \cite{mulas06a}. Recently, \cite{iglesias08} claimed a tentative identification of ionised naphthalene (C$_{10}$H$_{8}^+$) by the correspondence of three bands from its electronic spectrum with  three observed DIBs. 

\cite{mulas06b} showed that, in principle, it is possible to identify specific interstellar PAHs by the detection of the ro-vibrational emission bands that arise in the far-IR during the cooling cascade following UV excitation. The observation of these bands in the far-IR domain however requires airborne and satellite instruments due to strong atmospheric absorption, and will be one of the goals of the Herschel Space Observatory (HSO)\footnote{http://herschel.esac.esa.int/}. 
On the other hand, the rotational transitions of these
 molecules fall in the mm domain, and are readily accessible with ground-based  radio telescopes. 

\begin{figure}	
\begin{center}
\includegraphics[trim=3cm 3cm 3cm 3cm, clip,
width=3.8cm]{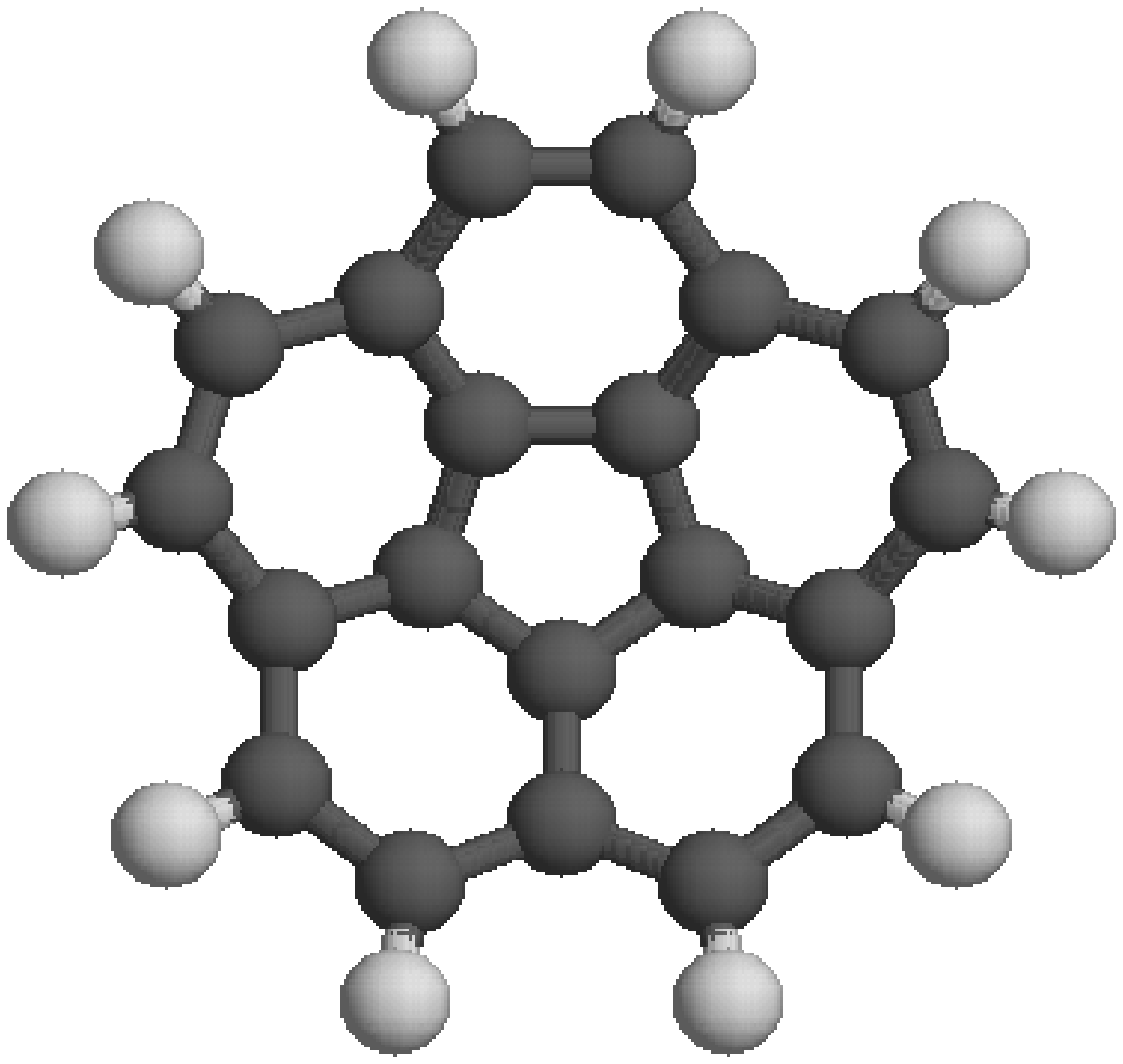}
\includegraphics[trim=3cm 3cm 3cm 3cm,
clip,width=3.8cm]{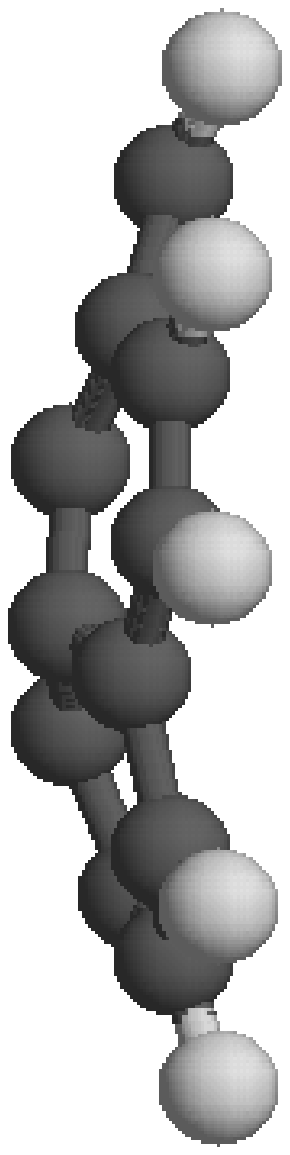}
\end{center}
\caption{Structure of the corannulene molecule (C$_{20}$H$_{10}$). The central pentagonal carbon ring determines the bowl-shaped structure of the molecule, with a permanent dipole moment of 2.07~D along the symmetry axis. }
\label{fig:structure}
\end{figure}

In a PAH, the absorption of a UV photon generally leads to fast  internal conversion of its energy into vibrational energy of the electronic ground state. This energy is then released by ro-vibrational emission in the mid-IR and far-IR ranges \citep[cf. e.g.,  models by][]{joblin02,mulas06c}. The intensity of the rotational spectrum scales with the square of the dipole moment but, unfortunately, most common neutral PAHs present very low (or zero) permanent dipole moments. Still there are a few exceptions \citep{lovas05,thorwirth07}, and  \cite{lovas05} proposed that a good PAH candidate for radio identification is corannulene (C$_{20}$H$_{10}$), which has a large dipole moment of 2.07~D (see Fig. \ref{fig:structure}). 
  \cite{thaddeus06} conducted a first search for corannulene in the molecular cloud TMC-1 searching for the low-J transitions reported in \cite{lovas05}. One difficulty with such observational strategy is that free PAHs are expected to be present at the surface of molecular clouds \citep{boulanger90, rapacioli05, berne07}. In these regions, the molecules are excited by UV photons and their rotational spectrum will differ from that of cold molecules excited by collisions \citep{rouan92}. 

There are several motivations to put further effort into the search for corannulene in space. It is a member  of the PAH population and its detection would provide the first firm evidence for the presence of such species in space.

Experimental studies \citep{lafleur93} indicate that corannulene is not a peculiar PAH and that it should not be classified as an unlikely component in natural mixtures. It is indeed expected to constitute a sizeable fraction of a mixture of small PAHs produced by pyrolysis of hydrocarbons 
which is commonly considered to be the formation pathway of PAHs in C\textendash rich outflows \citep{frenklach89, cherchneff92, cernicharo01}. In short, corannulene should be a good tracer of the small PAH population.

Furthermore, with its bowl-shaped geometry, it is also representative of the transition between planar PAHs and curved fullerenes, and several authors have proposed chemical pathways for the formation of C$_{60}$ involving corannulene \citep{haymet86, kroto88, chang92}. The prototype fullerene molecule, C$_{60}$, has been proposed in its cationic form (C$_{60}^+$) to account for at least two DIBs in the near-IR  \citep{foing94, galazutdinov00}.

In this paper we discuss  the concerted efforts between modelling and laboratory work that led us to perform a first observational campaign to search for corannulene in UV-irradiated environments. The Red Rectangle (RR) nebula was chosen since it is the brightest source in the AIBs in the sky and because it exhibits emission
features at nearly the same wavelengths as some DIBs \citep{schmidt80,scarrott92, sarre95, vanwinckel02}, making it a good source for validating the PAH model. Modelling and laboratory work are described in Secs. \ref{sec:modelling} and \ref{sec:laboratory}.  Observations and data reduction are presented in Sec. \ref{sec:observations}, and discussion is provided in Sec. \ref{sec:discussion}.

\section{Modelling}	\label{sec:modelling}
The emission model for a generic interstellar PAH molecule described in \cite{mulas98} has been extended and applied to individual PAHs in \cite{mulas06c}, where the far-IR spectra for a large sample of PAH species were presented. 
This model is applied here to calculate the rotational emission spectrum of a  C$_{20}$H$_{10}$ molecule in the radiation field of the RR halo, as defined in \cite{mulas06a}, i.e.,   the region of the RR nebula which is out of the bipolar cone and out of the dust torus surrounding the central binary system.
The model  input parameters are the UV-visible absorption cross-section, the vibrational modes and their Einstein A coefficients, the rotational constants and the dipole moment. 
The rotational constants were taken from \citet{lovas05}, while the absorption cross section, the frequencies and the A coefficients of the IR active modes were obtained by state-of-the-art quantum-mechanical calculations \citep[available in the PAH spectral database \textit{http://astrochemistry.ca.astro.it/database, }][]{malloci07}. For all other relevant molecular parameters, we used the same assumptions as in \citet{mulas06a}.

\begin{figure*}	
\begin{center}
\includegraphics[width=0.9\hsize]{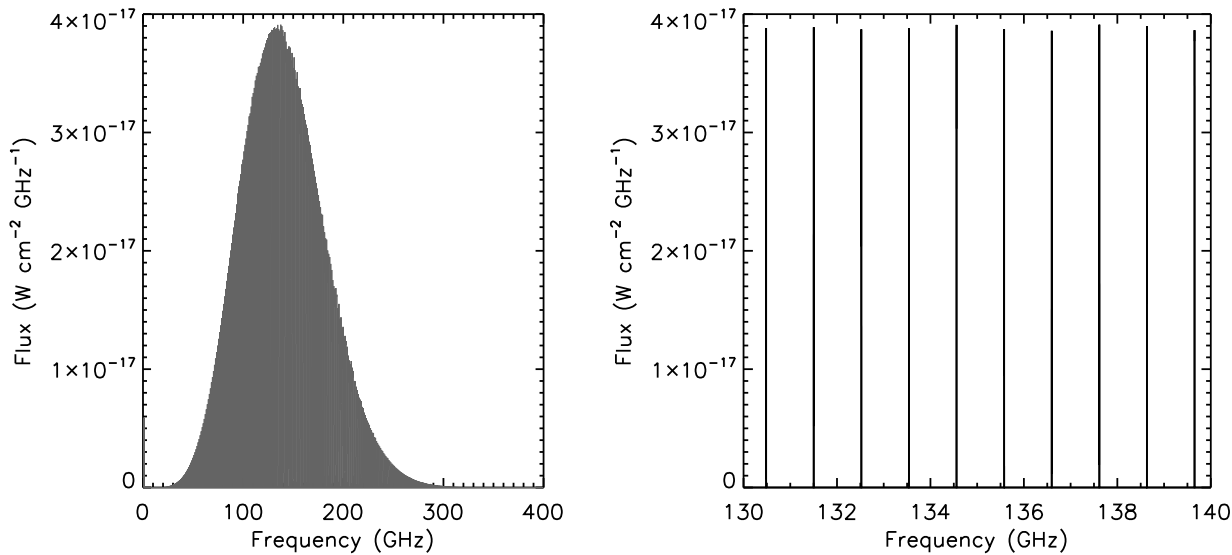}
\end{center}
\caption{The rotational emission spectrum of corannulene in the Red Rectangle calculated with the Monte-Carlo model peaks around 150~GHz (left panel) and is spread over a few hundreds GHz, with a spacing of about 1~GHz as shown by the zoom of a small region around the maximum (right panel). In the figure, a FWHM of 1~km~s$^{-1}$ has been assumed for the lines. The plot refers to the emission of corannulene molecules situated in the halo of the Red Rectangle, and has been normalised assuming that all the AIB flux measured by ISO-SWS was due to corannulene. }
\label{fig:theospec}
\end{figure*}

Since calculations are for a single C$_{20}$H$_{10}$ molecule, the synthetic spectrum is then scaled in the following way to allow the comparison with the observations:
 
\begin{itemize}
\item any PAH (including corannulene, if present) absorbs energy in the UV proportionally to its UV absorption  cross-section, which scales with the number of carbon atoms in the molecule \citep{joblin92}.
\item After the absorption of a UV photon, most of this energy is re-emitted in the AIBs. We integrated the AIB flux between 3 and 15~$\mu$m measured with the Short Wavelength Spectrograph (SWS) onboard the Infrared Space Observatory (ISO)   \citep{waters98}  to estimate the energy absorbed by all PAHs in the nebula. 
\item We calculated the  expected integrated flux emitted by a single corannulene molecule between 3 and 15~$\mu$m, and derived the normalisation factor to the flux measured by SWS.
\item We applied the same normalisation factor to the whole spectrum, including the rotational transitions.  
\end{itemize}

The Einstein A coefficients for spontaneous emission in the IR bands are typically between $10^{-2}$ and $10^{2}$ s$^{-1}$. In case of pure rotational transitions, typical A values are of the order of $10^{-5}$ -- $10^{-7}$ s$^{-1}$. Therefore, the IR and rotational emissions are expected to occur at different time scales. However, the IR cascade can bring the molecule in to a vibrationally excited (metastable) state, where only IR inactive modes are populated. The molecule can therefore spend a non-negligible amount of time in states like this, since their main relaxation channel is via a forbidden vibrational transition. With the modelling parameters
adopted, corannulene is estimated to spend roughly $60$\% of its time in the ground vibrational state,  the remaining 40\% being divided in a large number of different metastable  states.
A fraction of the pure rotational emission of the molecule occurs then from such states, in which rotational constants are slightly different from those of the ground vibrational state. Rotational lines from such metastable vibrational states are displaced from those emitted from the ground vibrational state, and do not contribute to the observed line intensity.
The estimated time spent in the metastable state depends on the assumed radiation field, and therefore on the spatial distribution of corannulene within the nebula, which we assumed to be in the halo similarly to the observed 11.3~$\mu$m emission \citep{waters98}.  If it is instead located more similarly to the 3.3~$\mu$m emission, which is closer to the central source, the photon absorption rate would be higher, the fraction of time spent in the ground state lower. The reverse is true if corannulene is located  further away from the source than we assumed. The fraction of time spent in the ground vibrational state depends also on our assumptions for the Einstein coefficients for the IR\textendash inactive modes \citep{mulas06c}. If corannulene turned out to have very unusual intensities for electric quadrupole vibrational transitions, with respect to IR\textendash active ones, this would correspondingly change the estimated lifetimes of the metastable states.

The associated emission in the rotational levels is shown in Fig. \ref{fig:theospec}, where a FWHM of 1~km~s$^{-1}$ is assumed and no scaling factor was applied to account for metastable states.  
The resulting rotational spectrum is expected to have its most intense lines around 150~GHz and to be spread over a few hundreds GHz with a line spacing of about 1 GHz.

\section{Laboratory work}\label{sec:laboratory}
Corannulene is a polar, symmetric-top PAH with a bowl shaped structure (see Fig. \ref{fig:structure}). The rotational energy levels of symmetric-top molecules are segregated into series of K-stacks, distinguished by the value of the rotational angular momentum K along the molecular symmetry axis. For parallel bands, the selection rule for the K quantum number is $\Delta K = 0$, so that only levels within the same stack can be connected through a rotational transition. In an ideal rigid symmetric top molecule, the J $\rightarrow$ J+1 transition has the same frequency along each stack, but in reality centrifugal distortion generally separates the transitions in well resolved lines, the high-J transitions being more affected by this effect. 
The high resolution rotational spectrum of corannulene has been measured by \cite{lovas05} by Fourier transform microwave spectroscopy  (FTMW) up to J = 19  with no K-splitting observed up to this line, leading to an upper limit of $\Delta_{JK} = 2.3\times10^{-6}$ for the centrifugal distortion constant.
For a radio astronomical detection, higher J transitions ($J > 100$) are required because of the expected excitation pattern of corannulene by the impinging UV radiation field. It is thus of great importance to know whether K-splitting is still negligible for high-J transitions. 

\begin{figure}	
\begin{center}
\includegraphics[width=\hsize]{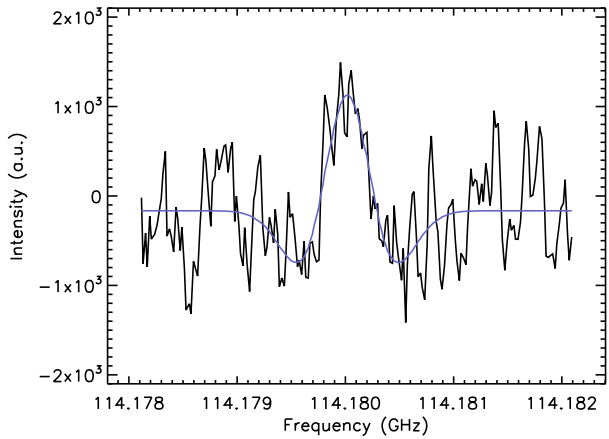}
\end{center}
\caption{The line detection at 114.18 GHz measured in the laboratory with the OROTRON jet spectrometer, assigned to the J = 112 $\leftarrow$ 111 transition of corannulene.}
\label{fig:labwork}
\end{figure}

Our laboratory measurements were performed using the Intracavity OROTRON jet spectrometer in Cologne, which is characterised by a frequency range of 112-156~GHz and a frequency resolution of 10-15~kHz \citep{surin01}. In this setup both the millimeter wave generator OROTRON and the supersonic jet apparatus are placed inside a vacuum chamber.  Corannulene was prepared by solution phase methods \citep{sygula01} and purified by chromatography on silica gel. The sample was crystalline powder  which was evaporated at a temperature of about 170$^\circ$ C, and  injected into the OROTRON cavity via a heated pinhole nozzle with Helium as carrier gas. A backing pressure of about 1~bar and a jet repetition rate of 5-10~Hz was used. At 114.18~GHz a  weak spectral feature was recorded after 10 minutes of integration time, which has been assigned to the  J = 112 $\leftarrow$ 111 transition of corannulene (see Fig. \ref{fig:labwork}). 
The central frequency is in agreement with predictions  based on the spectroscopic constants obtained from the FTMW measurements to  within 250 kHz. The line shows no K-splitting from which can be concluded that the corannulene molecule is very  rigid. Therefore, the intensities of different K-stack transitions sum up, enhancing the total intensity of each J $\rightarrow$ J+1 line by a factor of some tens, depending on the temperature of the gas  \citep{thaddeus06}. Combining the effects of polarity and the stiffness of the molecule, the intensity of the radio spectrum of corannulene is expected to be 3-4 orders of magnitude stronger than that of a typical polar PAH of the same size and abundance, making corannulene an excellent candidate for radio astronomical detection. 
The rotational constants derived from the laboratory work have been used to refine the band positions given in the synthetic spectrum. 
For the measurement of the corannulene line described in this paper 30 mg of corannulene had to be spent. 
Further laboratory measurements in the mm-waveband are in progress and will be described in more detail elsewhere (Herberth, Giesen, in preparation).

\section{Source description and observations}	\label{sec:observations}
The Red Rectangle (RR) is a biconical C-rich nebula which surrounds a post-AGB binary system composed of the primary star HD 44179 (spectral type A0) and its luminous giant companion. It has been observed in the mid-IR with the SWS instrument onboard ISO and with the Infrared Spectrograph (IRS) onboard the Spitzer Space Telescope. The dimensions on the plane of the sky of the infrared bipolar nebula at 11.3 $\mu$m are about 10" x 20" \citep{waters98}. A warm, high density disk ($n_H > 3~10^4$~cm$^{-3}$, 30~K $<$ T $<$ 400~K)  traced by  CO emission \citep{bujarrabal05} surrounds the binary star with a thickness of roughly 3" along the symmetry axis.  

\begin{table*}	
\centering
\begin{tabular}{l|cccccccc}
Transition				&	Frequency 	&	Beam size	&	$B_{eff}$		&	T$_{sys}$		&	$\sigma_{mb}$  &	    Bandwidth	&	F$_{3\sigma}$		&	F$_{model}$		
\\
 (J+1 $\rightarrow$ J)	&	(GHz)		&	(")			&				&	(K)			&	(mK)			 &		 (MHz)	&	($10^{-25}$~W~cm$^{-2}$)			&	($10^{-21}$~W~cm$^{-2}$)\\
\hline
\hline
84 $\rightarrow$ 83				&	85.643			&	29			&	0.78			&	117			&	10	&		80	&	1.4	&	5.7		\\
86 $\rightarrow$ 85				&	87.682			&	28			&	0.77			&	107			&	12	&		80	&	1.7	&	6.2		\\
108 $\rightarrow$ 107			&	110.104			&	22			&	0.75			&	170			&	14	&		80	&	1.6	&	13		\\
111 $\rightarrow$ 110			&	113.316			&	22			&	0.74			&	276			&	21	&		120	&	2.3	&	14		\\
135 $\rightarrow$ 134			&	137.615			&	17			&	0.70			&	219			&	8	&		80	&	0.9	&	19		\\
215 $\rightarrow$ 214			&	219.059			&	11			&	0.55			&	346			&	45	&		80	&	5.4	&	5.3		\\
216 $\rightarrow$ 215			&	220.076			&	11			&	0.54			&	328			&	20	&		80	&	2.4	&	5.2		\\
223 $\rightarrow$ 222			&	227.197			&	11			&	0.53			&	269			&	23	&		80	&	2.9	&	4.0		\\
226 $\rightarrow$ 225			&	230.248			&	11			&	0.52			&	389			&	30	&		80	&	3.8	&	3.7		\\
238 $\rightarrow$ 237			&	242.450			&	10			&	0.50 			&	525			&	22	&		80	&	2.8	&	2.3		\\
257 $\rightarrow$ 256			&	261.763			&	8			&	0.46			&	546			&	39	&		40	&	5.4	&	0.9		\\
\hline
\end{tabular}
\caption{Summary of the observations towards the RR in different frequency ranges, corresponding to the expected rotational transitions of corannulene. On average, the 3~mm observations have a lower rms, but suffer more from beam dilution effects compared to the 1~mm observations. }
\label{tab:observations}
\end{table*}

\begin{table*}	
\begin{tabular}{l|cccccc}
Transition				&	Frequency 	&	Beam size	&	$B_{eff}$		&	T$_{sys}$		&	Area  &	    Bandwidth\\
 (J+1 $\rightarrow$ J)	&	(GHz)		&	(")			&				&	(K)			&	(K~km~s$^{-1}$)			 &		 (MHz)	\\
\hline
\hline
$^{12}$CO (2 $\rightarrow $ 1)		&	230.538			&	10			&	0.52			&	486			& 6.8		&		40			\\
$^{13}$CO (1 $\rightarrow $ 0)		&	110.201			&	22			&	0.75			&	159			& 0.3		&		40		\\
$^{13}$CO (2 $\rightarrow $ 1)		&	220.398			&	11			&	0.55			&	339			& 1.7 	&		40				\\

\hline
\end{tabular}
\caption{Summary of $^{12}$CO and $^{13}$CO observations towards the Red Rectangle.  }
\label{tab:co}
\end{table*}

Observations were performed with the IRAM 30m telescope in Pico Veleta in February 2008.  
The observed frequency ranges cover several high-J transitions of C$_{20}$H$_{10}$ at 1, 2 and 3 mm. Weather conditions were acceptable for most of the time. 
We used the wobbling secondary observing mode,  with a beam separation of $\pm$100" in
azimuth. This observing mode ensures stable and flat baselines and is well
adapted to compact sources such as the Red Rectangle.

Pointing was made on the central binary star HD44179 ($\alpha_{2000}$: 06:19:58.216, $\delta_{2000}$: -10:38:14:691).
Mars and Orion ($\alpha_{2000}$: 05:35:14.5, $\delta_{2000}$: -05:22:30.00) were used as reference sources for calibration and pointing. Pointing was accurate within 3" during all  observations. The flexibility of the four receivers and of the VESPA correlator allowed to cover several frequency ranges within one configuration, with a spectral resolution of 40~kHz and a total band pass of at least 40~MHz. We also used the  1~MHz resolution filterbanks to obtain  broadband spectra (250 MHz) with lower spectral resolution.
 During all  observations,  one of the backends was dedicated to the $^{13}$CO~(1-0), (2-1), or  $^{12}$CO(2-1) transition, to check whether the telescope was pointed on 
source, and to monitor the calibration accuracy.  The observed $^{12}$CO intensity and line width  are in good agreement  with previous observations of the Red Rectangle with the IRAM 30m telescope reported in \cite{jura95}. The summary of the $^{12}$CO and $^{13}$CO observations are reported in Table \ref{tab:co}. 

\begin{figure}	
\begin{center}
\includegraphics[trim=2cm 0cm 0.5cm 2.5cm , clip,  angle=-90, width=0.48\hsize]{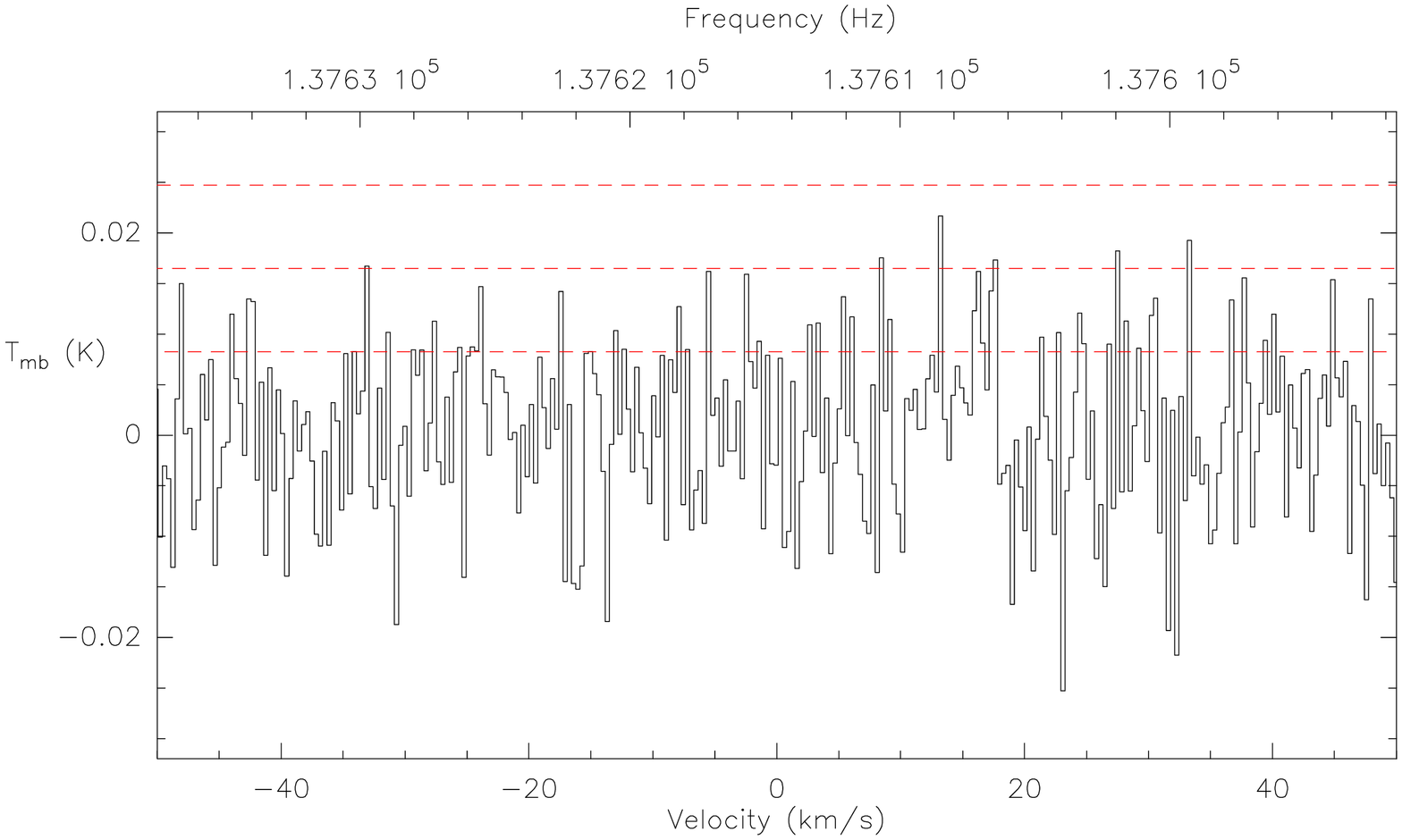}
\includegraphics[trim=2cm 0cm 0.5cm 2.5cm, clip ,  angle=-90, width=0.48\hsize]{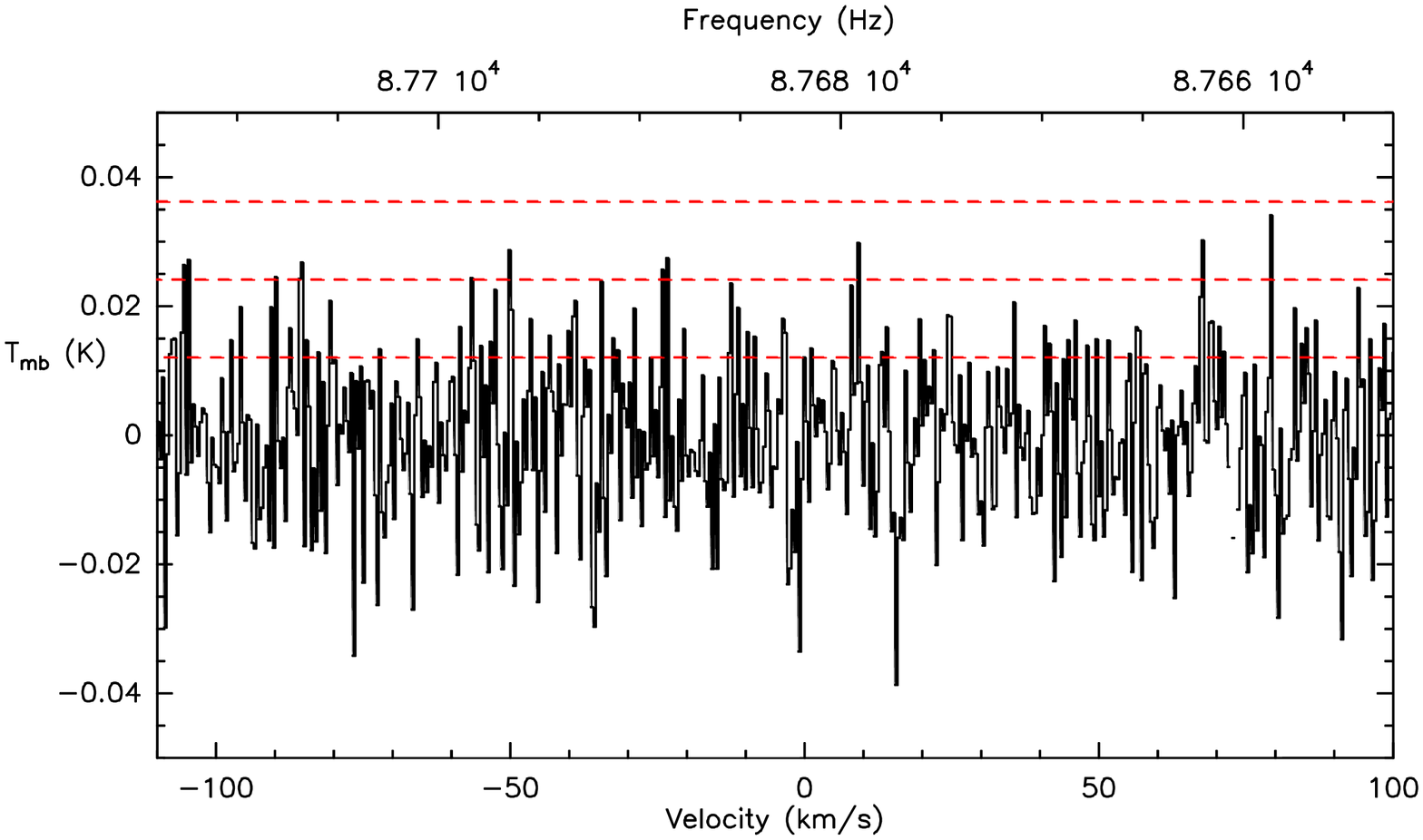}
\includegraphics[trim=2cm 0cm 0.5cm 2.5cm, clip ,  angle=-90, width=0.48\hsize]{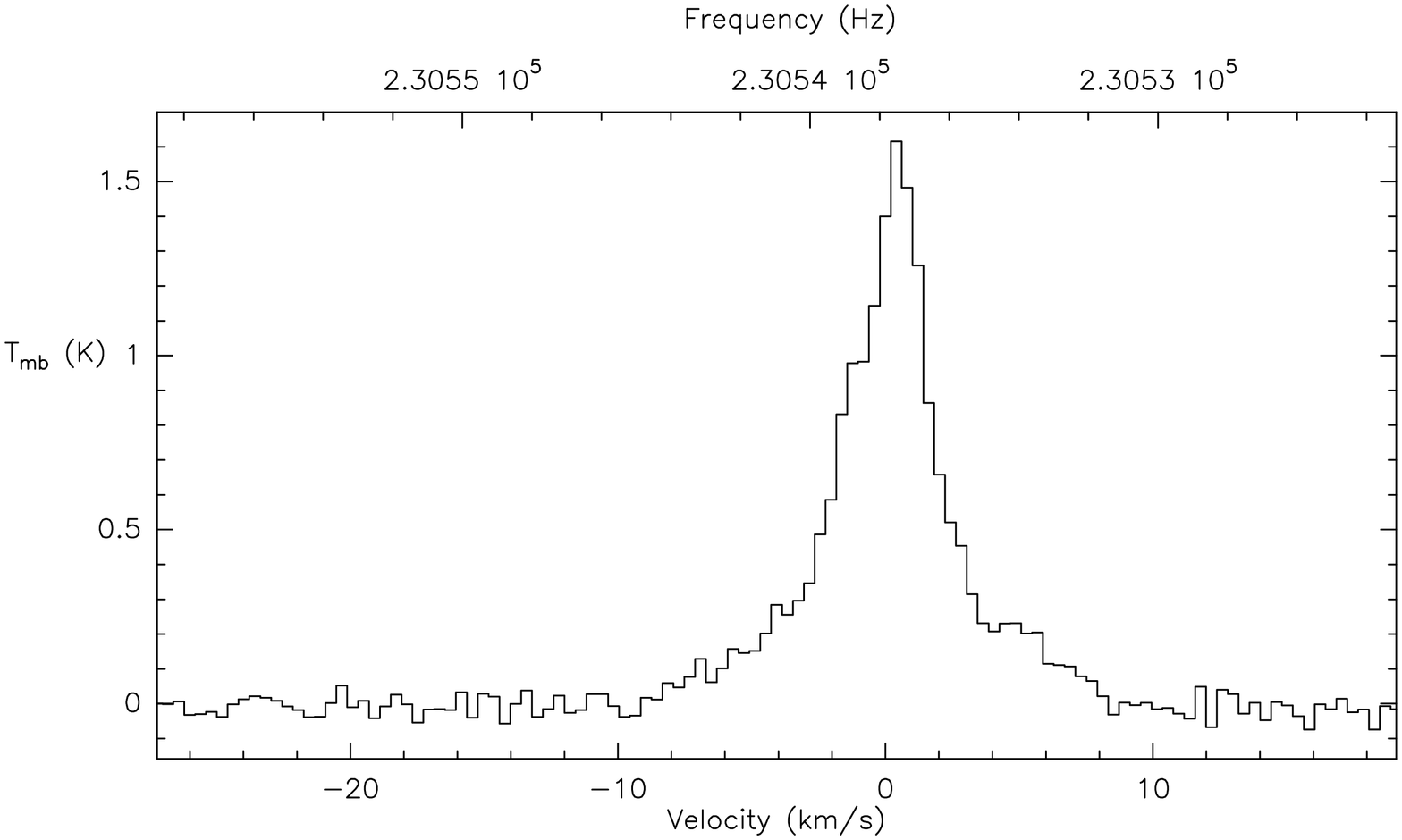}
\includegraphics[trim=2cm 0cm 0.5cm 2.5cm, clip ,  angle=-90, width=0.48\hsize]{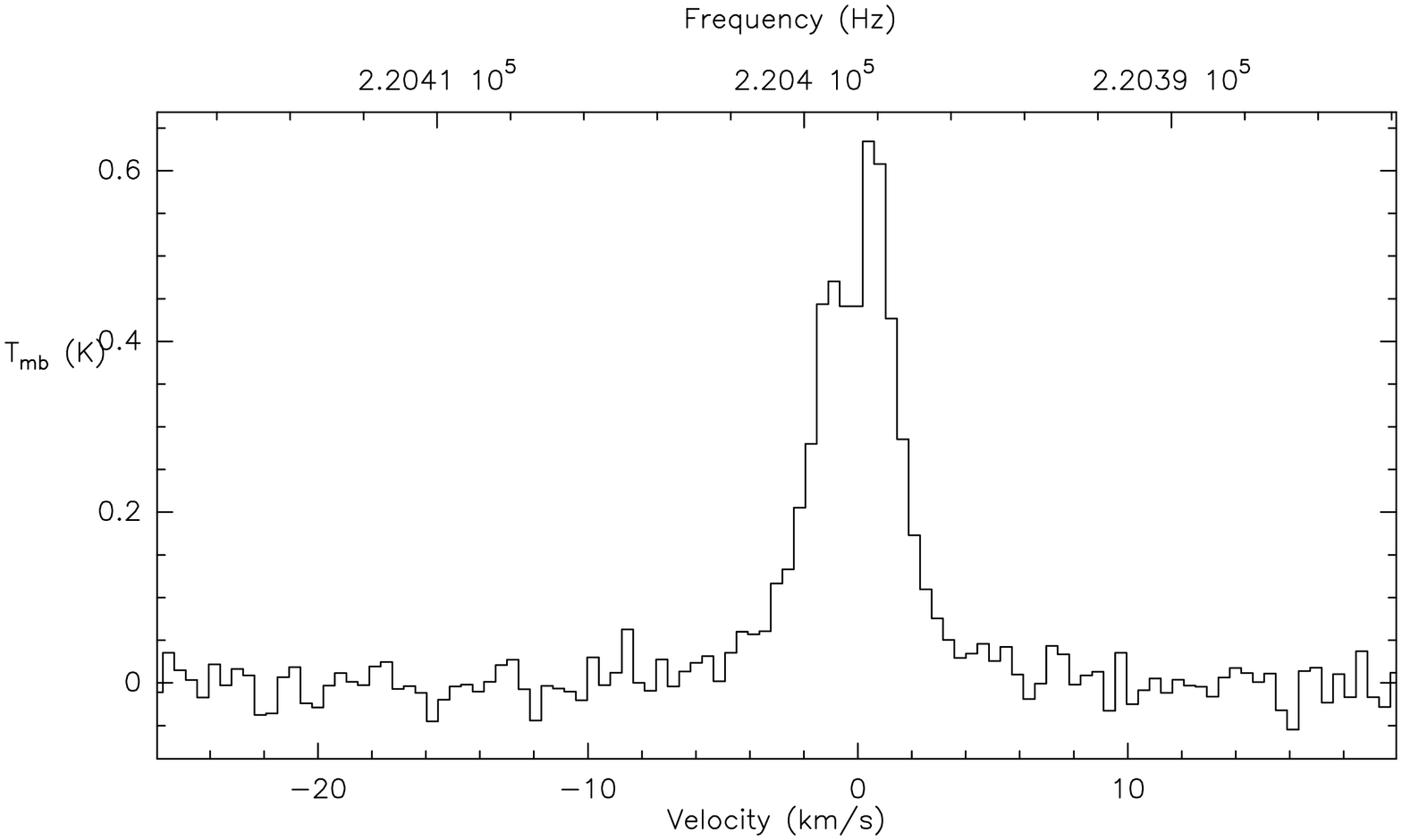}
\includegraphics[trim=2cm 0cm 0.5cm 2.5cm, clip ,  angle=-90, width=0.48\hsize]{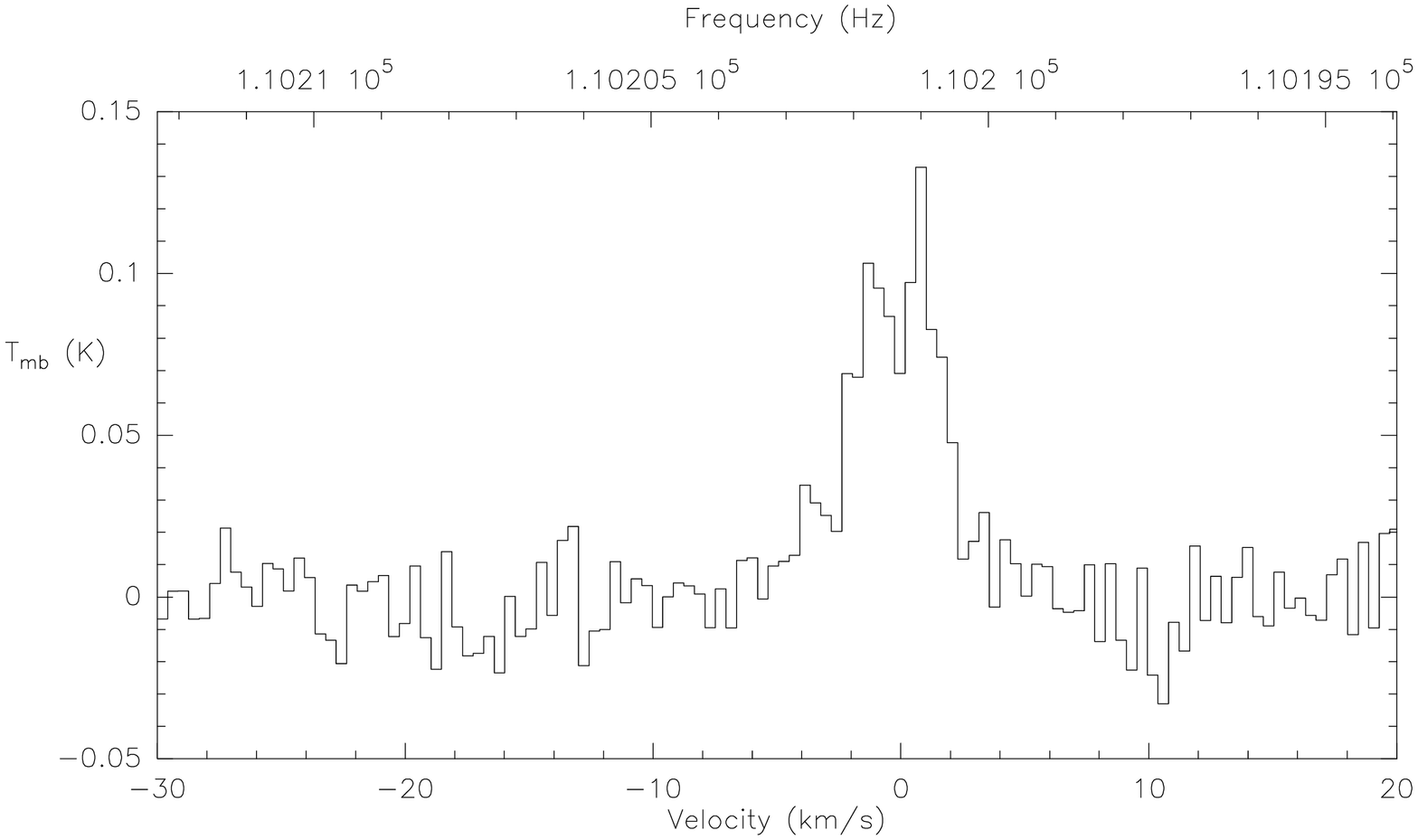}
\end{center}
\caption{Observations of the expected frequency range for the 135 $\rightarrow$ 134 transition at 137 GHz, the 86 $\rightarrow$ 85 transition at 88 GHz of C$_{20}$H$_{10}$, the $^{12}$CO(1-0) and $^{13}$CO(1-0),(2-1) transitions. Horizontal red lines indicate the first 3 sigma levels. See the electronic version of the article for further plots. }
\label{fig:spectra}
\end{figure}

To improve the quality of the observed spectra, we manually cut isolated spikes out of 4$\sigma$ (the two nearest channels being lower than 3$\sigma$), and discarded observations with anomalous system temperature or with a high sky opacity. Different observations of the same transitions were averaged, and the antenna temperature was scaled with the telescope main beam efficiency by the relation $T_{mb}=T^*_A/\eta_{mb}$, where $\eta_{mb}=F_{eff}/B_{eff}$. The central frequency, beam width, beam efficiency, system temperature, rms level and bandwidth are reported in Table \ref{tab:observations} for each of the observed frequency ranges. The expected flux derived from the synthetic spectrum and the 3$\sigma$ detection limit  corrected for beam dilution are also reported for the corannulene observations.  The rms level was calculated after smoothing the spectra to the velocity resolution of 0.4~km~s$^{-1}$. In Fig. \ref{fig:spectra} we show the observed spectra in the regions where the 135 $\rightarrow$ 134  and 86 $\rightarrow$ 85 transitions of C$_{20}$H$_{10}$ are expected, and the $^{12}$CO and $^{13}$CO  observations (further plots can be found in the electronic version of the article).  We did not detect any corannulene line at any frequency. The best rms level for corannulene has been obtained for the transitions J+1 $\rightarrow$ J = 135 $\rightarrow$ 134, 84$\rightarrow$ 83 and  86$\rightarrow$ 85.

\section{Results and discussion}	\label{sec:discussion}
To calculate an upper limit to the abundance of corannulene in the source, we assumed a 3$\sigma$ detection limit and a gaussian profile with a FWHM of 1 km~s$^{-1}$. 
Before comparing the observed rms to the model, we have to take into account the effect of beam dilution. As a first approximation, we assume that PAH emission is distributed homogeneously in a  20"x10" area  as observed for AIB emission \citep{waters98}. 

The best noise levels are reached, on average, with the 3~mm observations, but the 1 and 2~mm observations suffer less from beam dilution.  To compute a reliable upper limit to the abundance of corannulene, we chose the 135 $\rightarrow$ 134 transition at 2~mm, which is not only the transition with the lowest rms, but also the best compromise between the loss of flux due to beam dilution and  expected intensity derived from  the synthetic spectrum. 

The synthetic spectrum has been normalised as if all the AIB flux was due to corannulene, as explained in Sect. \ref{sec:modelling}. The ratio between observations and theoretical predictions can be turned into  an upper limit for the fraction of AIB flux due to corannulene, which corresponds to the fractional abundance of carbon  locked up into corannulene relative to the total abundance of carbon in PAHs. 
The synthetic spectrum predicts for the 135 $\rightarrow$ 134 transition an integrated flux of $F_{model} =1.9\times10^{-20}$~W~cm$^{-2}$. We compare this value with the area of the 3$\sigma$ gaussian for the same transition, corrected for beam dilution: F$_{3\sigma} = 9.0\times10^{-26}$~W~cm$^{-2}$. 

An upper limit for the fraction of carbon in C$_{20}$H$_{10}$ compared to the total abundance of carbon locked in PAHs can be obtained by the ratio F$_{3\sigma}$/F$_{model}$. This ratio has to be modified to take into account the fact that emitting C$_{20}$H$_{10}$ is not always in its ground state. An average factor of 0.6 was derived in Sect. \ref{sec:modelling}. 
Also,  the resulting estimate applies for the C$_{20}$H$_{10}$ main isotope. 
Different isotopologues have slightly different rotational constants, which  have rotational lines in displaced positions. Assuming standard solar system isotopic ratios  and no fractionation, about  25\% of corannulene molecules are expected to contain one or more $^{13}$C atoms which results in an additional correction. Substitutions of $^1$H with $^2$D occur in a negligible fraction of the molecules.   
 In a circumstellar envelope around a post-AGB star, isotopic ratios depend on the detailed history of the last phases of its precursor, and can be significantly different from the accepted solar values. For the RR, recent observations however failed to detect  isotopomers of C$_2$, CN and CH containing $^{13}$C, setting a lower limit of 22 for the $^{12}$C/$^{13}$C ratio in the RR nebula \citep{bakker97}. Assuming this ratio, instead of the standard value of 89, would result in a factor of 1.7 for our final lower limit on the abundance and column density of corannulene. Given that the $^{12}$C/$^{13}$C ratio in the RR must be $\ge$ 22, this does not significantly alter our conclusions.

With these assumptions, and solar $^{12}$C$/^{13}$C ratio, we derive a value of  $1.0 \times10^{-5}$ as our best limit for the fraction of carbon in C$_{20}$H$_{10}$ relative to the total carbon in all PAHs. Assuming that $\sim$ 20\% of total carbon atoms are locked in PAHs \citep{joblin92, tielens05}, this turns into a value of $2.0 \times10^{-6}$ of total carbon in corannulene. This result is much more stringent than the limit derived by \cite{thaddeus06} for corannulene in TMC-1,  $1\times10^{-5}$.

This can be compared with the models of the size distributions of interstellar 
PAHs in the literature. With $a$ the PAH radius, \cite{desert90} used a size 
distribution $N_{PAH}(a) \propto a^{-2}$ whereas \citet{draine98} proposed a 
lognormal distribution. Assuming that the fraction of carbon  locked in PAHs 
relative to hydrogen is $6\times10^{-5}$ \citep{joblin92, li01}, we can use 
the derived upper limit to constrain the corannulene abundance (relative to 
hydrogen) to be N$_{C_{20}H_{10}}/$N$_H < 3\times10^{-11}$.

\begin{figure}
\begin{center}
\includegraphics[width=\hsize]{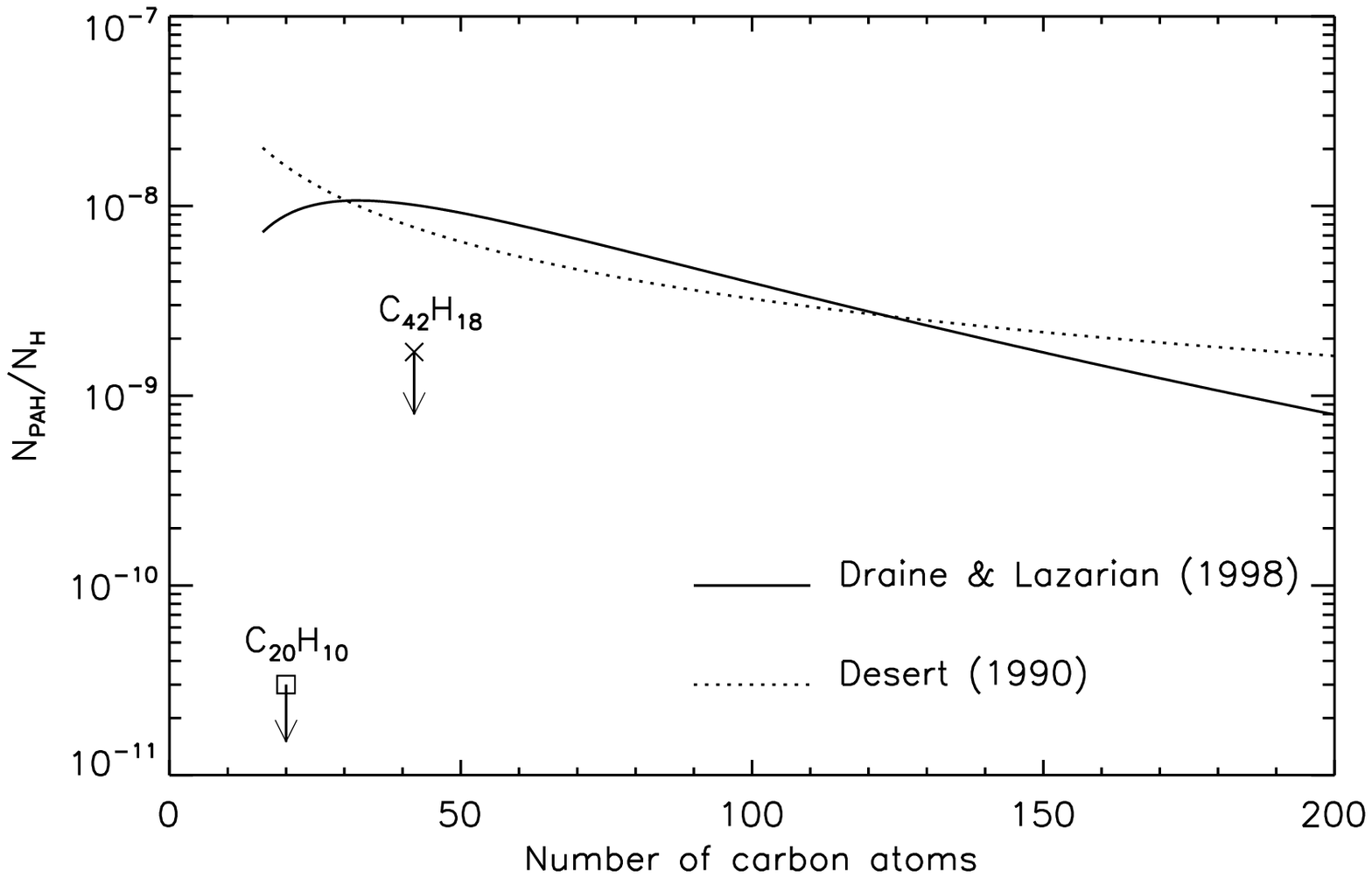}
\end{center}
\caption{Size distributions of PAHs according to 
\citet{desert90} and \citet{draine98}, normalised to a carbon abundance in 
PAHs of $6\times10^{-5}$ relative to hydrogen \citep{joblin92, draine98}. 
The two upper limits reported refer to the observations of corannulene in 
the RR (this paper) and of C$_{42}$H$_{18}$  in the diffuse interstellar 
medium \citep{kokkin08}. The discrepancy between  models and observations 
supports the photo-destruction of small PAHs in space.}
\label{fig:sizeabund}
\end{figure}

Figure \ref{fig:sizeabund} shows both distributions using the relation 
$a=0.9\sqrt{N_C}$\AA\  for compact molecules \citep{omont86} and normalising 
the carbon content in PAHs to  $6\times10^{-5}$ relative to hydrogen. In the 
figure, the distributions are represented as histograms with a bin size of 1 
in N$_C$, to be directly comparable with the upper limits derived for 
corannulene (this work) and  for the  larger species  C$_{42}$H$_{18}$ that 
has been measured in the visible range for the diffuse interstellar medium 
\citep{kokkin08}. It is difficult to assess the significance of such a 
straight comparison, however, since on one side we have a specific molecule 
and its abundance, on the other side we have the summed abundance of all PAHs 
in the same size bin. This would involve considering all isomers with the 
same chemical formula, and assuming some relative abundances among them.  
In the absence of more specific information, we will use the experimental 
results cited in Sect.~\ref{sec:introduction} about the minimum mass fraction 
of corannulene that is always formed in pyrolysis experiments. In the 
chromatograms of the aromatic species produced in such flames, a few tens 
of species appear to constitute most of the mixture of molecules with less 
than 100~C atoms,  fewer than ten of them being by far more abundant 
\citep{lafleur93}.

Using the PAH spectral database \citep{malloci07} we determined that, for neutral PAHs, there is a tight linear correlation (correlation coefficient $r = 0.94$) between the number of carbon atoms and the absorbance integrated in the (236-500) nm range used by the chromatogram in Figure 3 of \cite{lafleur93}. This means
that we can use these integrated absorbances to quantify N$_C$ for each identified PAH in that chromatogram. We then derived the mass fraction of pyrene compared to all PAHs, and that of corannulene, assuming it corresponds to 1\% of the pyrene mass, which is the lowest value reported by \cite{lafleur93}.
This leads to a lower limit of $\sim 4 \times 10^{-4}$ for the fraction in mass of corannulene in the mixtures of small PAHs produced in such experiments.
Together with our observational upper limit on the fraction of carbon atoms in corannulene with respect to the total in all PAHs, this yields an upper limit of $\sim$2\% for the mass fraction of small PAHs with respect to the total  (3.4\% assuming the lower limit for $^{12}$C$/^{13}$C).
Integrating the two size distributions (see Fig. \ref{fig:sizeabund}) in the small molecule range (N$_C\leq 50$) leads to a fraction of carbon in small PAHs vs the total of $\sim$19\%, an order of magnitude larger than our upper limit. 
Based on the general assumption that the formation of PAHs in evolved stars can be described by flame chemistry (cf. Sect. Introduction), we can conclude that small PAHs in the RR are significantly less abundant than the values predicted by the size distributions reported in Fig. \ref{fig:sizeabund}.

The under-abundance of small PAHs in the RR can be due either to selective destruction of small-sized PAHs  \citep{lepage03, allain96b, allain96a} or to the fact that PAHs grew to larger sizes directly in the C-rich outflow of the RR nebula progenitor where they were originally synthesised. Interestingly, the analysis of observations of the 3.3~$\mu$m AIB and its overtone at 1.68~$\mu$m in the young planetary nebula IRAS 21282+5050 showed that PAHs emitting in these bands contain about 60 C \citep{geballe94}. This rises the question of the minimum size of PAHs that are formed in evolved stars, an important issue considering that these PAHs also appear to be the parental species of interstellar PAHs \citep{joblin08}.

To determine whether photodissociation can destroy small PAHs in the RR halo, we estimated the rate of the dominant photodissociation channel for corannulene, namely the loss of an H atom, as a function of the internal energy U. 
This dissociation rate $k_d(U)$ was calculated using the results on the coronene cation (C$_{24}$H$_{12}^+$) from the PIRENEA set-up \citep{joblin09}, and including corrections for the variation of the density of states between C$_{20}$H$_{10}$ and C$_{24}$H$_{12}^+$ (cf. formula (7) in \cite{boissel97}).
We combined this rate with the distribution of excitation energies of C$_{20}$H$_{10}$ in the RR halo obtained from the  Monte\textendash Carlo simulation discussed above. Multiplying the two and integrating, we obtained a photodissociation rate of $\sim 1.1\times 10^{-8}$~s$^{-1}$, corresponding to a lifetime of $\sim 2.9$~years.
The reaction rate of dehydrogenated corannulene with H can be estimated at a value of at least $2\times10^{-5}$~s$^{-1}$, using a rate coefficient of $\sim2\times10^{-10}$~cm$^3$~s$^{-1}$ \citep{lepage99} and a hydrogen density larger than 10$^5$~cm$^{-3}$ \citep{menshchikov02}. This is several orders of magnitudes larger than the dissociation rate. Therefore, corannulene is expected to survive in its neutral hydrogenated form in the RR nebula.

 From this work we can conclude that  radio spectroscopy can be a powerful tool to detect  specific polar PAHs. Our results in the RR show that it can be  also an efficient way to constrain the abundance of small PAHs.  Our analysis of these results suggests that the under-abundance of small PAHs in the RR is related to the production mechanism for these species and not to subsequent UV destruction. This conclusion will have to be further tested by both observations in other environments and by laboratory studies on the formation of PAHs in conditions that mimic the environment of evolved carbon stars. 

\section*{Acknowledgements}
 We thank the referee for his useful comments.  This work was supported by the European Research Training
Network ÒMolecular UniverseÓ (MRTN-CT-
2004-512302), and the CNRS PICS 4260 "Croissance et destruction des macromol\'ecules carbon\'ees interstellaires", which are acknowledged. P. Pilleri acknowledges also financial support from the Master \& Back program of the "Regione Autonoma della Sardegna". Financial support from the U.S. National Science Foundation (Department of Energy),  the Deutsche Forschungsgemeinschaft and the Land Niedersachsen are also gratefully acknowledged. G. Mulas and G. Malloci acknowledge financial support by MIUR under project CyberSar, call 1575/2004 of PON 2000-2006. 

\bibliographystyle{mn2e}
\bibliography{biblio}

\begin{thebibliography}{}

\bibitem[\protect\citeauthoryear{{Allain}, {Leach} \& {Sedlmayr}}{{Allain}
  et~al.}{1996a}]{allain96b}
{Allain} T.,  {Leach} S.,    {Sedlmayr} E.,  1996a, \aap, 305, 602

\bibitem[\protect\citeauthoryear{{Allain}, {Leach} \& {Sedlmayr}}{{Allain}
  et~al.}{1996b}]{allain96a}
{Allain} T.,  {Leach} S.,    {Sedlmayr} E.,  1996b, \aap, 305, 616

\bibitem[\protect\citeauthoryear{{Allamandola}, {Tielens} \&
  {Barker}}{{Allamandola} et~al.}{1985}]{allamandola85}
{Allamandola} L.~J.,  {Tielens} A.~G.~G.~M.,    {Barker} J.~R.,  1985, \apjl,
  290, L25

\bibitem[\protect\citeauthoryear{{Allamandola}, {Tielens} \&
  {Barker}}{{Allamandola} et~al.}{1989}]{allamandola89}
{Allamandola} L.~J.,  {Tielens} A.~G.~G.~M.,    {Barker} J.~R.,  1989, \apjs,
  71, 733

\bibitem[\protect\citeauthoryear{{Bakker}, {van Dishoeck}, {Waters} \&
  {Schoenmaker}}{{Bakker} et~al.}{1997}]{bakker97}
{Bakker} E.~J.,  {van Dishoeck} E.~F.,  {Waters} L.~B.~F.~M.,    {Schoenmaker}
  T.,  1997, \aap, 323, 469

\bibitem[\protect\citeauthoryear{{Bern{\'e}}, {Joblin}, {Deville}, {Smith},
  {Rapacioli}, {Bernard}, {Thomas}, {Reach} \& {Abergel}}{{Bern{\'e}}
  et~al.}{2007}]{berne07}
{Bern{\'e}} O.,  {Joblin} C.,  {Deville} Y.,  {Smith} J.~D.,  {Rapacioli} M.,
  {Bernard} J.~P.,  {Thomas} J.,  {Reach} W.,    {Abergel} A.,  2007, \aap,
  469, 575

\bibitem[\protect\citeauthoryear{{Boissel}, {de Parseval}, {Marty} \&
  {Lef{\`e}vre}}{{Boissel} et~al.}{1997}]{boissel97}
{Boissel} P.,  {de Parseval} P.,  {Marty} P.,    {Lef{\`e}vre} G.,  1997, J.
  Chem. Phys., 106, 4973

\bibitem[\protect\citeauthoryear{{Boulanger}, {Falgarone}, {Puget} \&
  {Helou}}{{Boulanger} et~al.}{1990}]{boulanger90}
{Boulanger} F.,  {Falgarone} E.,  {Puget} J.~L.,    {Helou} G.,  1990, \apj,
  364, 136

\bibitem[\protect\citeauthoryear{{Bujarrabal}, {Castro-Carrizo}, {Alcolea} \&
  {Neri}}{{Bujarrabal} et~al.}{2005}]{bujarrabal05}
{Bujarrabal} V.,  {Castro-Carrizo} A.,  {Alcolea} J.,    {Neri} R.,  2005,
  \aap, 441, 1031

\bibitem[\protect\citeauthoryear{{Cecchi-Pestellini}, {Malloci}, {Mulas},
  {Joblin} \& {Williams}}{{Cecchi-Pestellini}
  et~al.}{2008}]{cecchipestellini08}
{Cecchi-Pestellini} C.,  {Malloci} G.,  {Mulas} G.,  {Joblin} C.,    {Williams}
  D.~A.,  2008, \aap, 486, L25

\bibitem[\protect\citeauthoryear{{Cernicharo}, {Heras}, {Tielens}, {Pardo},
  {Herpin}, {Gu{\'e}lin} \& {Waters}}{{Cernicharo} et~al.}{2001}]{cernicharo01}
{Cernicharo} J.,  {Heras} A.~M.,  {Tielens} A.~G.~G.~M.,  {Pardo} J.~R.,
  {Herpin} F.,  {Gu{\'e}lin} M.,    {Waters} L.~B.~F.~M.,  2001, \apjl, 546,
  L123

\bibitem[\protect\citeauthoryear{{Chang}, {Naim}, {Ahmed}, {Goodloe} \&
  {Shevlin}}{{Chang} et~al.}{1992}]{chang92}
{Chang} T.-M.,  {Naim} A.,  {Ahmed} S.~N.,  {Goodloe} G.,    {Shevlin} P.~B.,
  1992, J. Am. Chem. Soc., 114, 7603

\bibitem[\protect\citeauthoryear{{Cherchneff}, {Barker} \&
  {Tielens}}{{Cherchneff} et~al.}{1992}]{cherchneff92}
{Cherchneff} I.,  {Barker} J.~R.,    {Tielens} A.~G.~G.~M.,  1992, \apj, 401,
  269

\bibitem[\protect\citeauthoryear{{D\'esert}, {Boulanger} \& {Puget}}{{D\'esert}
  et~al.}{1990}]{desert90}
{D\'esert} F.-X.,  {Boulanger} F.,    {Puget} J.~L.,  1990, \aap, 237, 215

\bibitem[\protect\citeauthoryear{{Donn}}{{Donn}}{1968}]{donn68}
{Donn} B.,  1968, \apjl, 152, L129+

\bibitem[\protect\citeauthoryear{{Draine} \& {Lazarian}}{{Draine} \&
  {Lazarian}}{1998}]{draine98}
{Draine} B.~T.,  {Lazarian} A.,  1998, \apj, 508, 157

\bibitem[\protect\citeauthoryear{{Foing} \& {Ehrenfreund}}{{Foing} \&
  {Ehrenfreund}}{1994}]{foing94}
{Foing} B.~H.,  {Ehrenfreund} P.,  1994, \nat, 369, 296

\bibitem[\protect\citeauthoryear{{Frenklach} \& {Feigelson}}{{Frenklach} \&
  {Feigelson}}{1989}]{frenklach89}
{Frenklach} M.,  {Feigelson} E.~D.,  1989, \apj, 341, 372

\bibitem[\protect\citeauthoryear{{Galazutdinov}, {Kre{\l}owski}, {Musaev},
  {Ehrenfreund} \& {Foing}}{{Galazutdinov} et~al.}{2000}]{galazutdinov00}
{Galazutdinov} G.~A.,  {Kre{\l}owski} J.,  {Musaev} F.~A.,  {Ehrenfreund} P.,
   {Foing} B.~H.,  2000, \mnras, 317, 750

\bibitem[\protect\citeauthoryear{{Geballe}, {Joblin}, {D'Hendecourt}, {Jourdain
  de Muizon}, {Tielens} \& {Leger}}{{Geballe} et~al.}{1994}]{geballe94}
{Geballe} T.~R.,  {Joblin} C.,  {D'Hendecourt} L.~B.,  {Jourdain de Muizon} M.,
   {Tielens} A.~G.~G.~M.,    {Leger} A.,  1994, \apjl, 434, L15

\bibitem[\protect\citeauthoryear{{Haymet}}{{Haymet}}{1986}]{haymet86}
{Haymet} A.~D.~J.,  1986, J. Am. Chem. Soc., 108, 319

\bibitem[\protect\citeauthoryear{{Iglesias-Groth}, {Manchado},
  {Garc{\'{\i}}a-Hern{\'a}ndez}, {Gonz{\'a}lez Hern{\'a}ndez} \&
  {Lambert}}{{Iglesias-Groth} et~al.}{2008}]{iglesias08}
{Iglesias-Groth} S.,  {Manchado} A.,  {Garc{\'{\i}}a-Hern{\'a}ndez} D.~A.,
  {Gonz{\'a}lez Hern{\'a}ndez} J.~I.,    {Lambert} D.~L.,  2008, \apjl, 685,
  L55

\bibitem[\protect\citeauthoryear{{Joblin}, {L\'eger} \& {Martin}}{{Joblin}
  et~al.}{1992}]{joblin92}
{Joblin} C.,  {L\'eger} A.,    {Martin} P.,  1992, \apjl, 393, L79

\bibitem[\protect\citeauthoryear{{Joblin}, {Szczerba}, {Bern{\'e}} \&
  {Szyszka}}{{Joblin} et~al.}{2008}]{joblin08}
{Joblin} C.,  {Szczerba} R.,  {Bern{\'e}} O.,    {Szyszka} C.,  2008, \aap,
  490, 189

\bibitem[\protect\citeauthoryear{{Joblin}, {Toublanc}, {Boissel} \&
  {Tielens}}{{Joblin} et~al.}{2002}]{joblin02}
{Joblin} C.,  {Toublanc} D.,  {Boissel} P.,    {Tielens} A.~G.~G.~M.,  2002,
  Mol. Phys., 100, 3595

\bibitem[\protect\citeauthoryear{{Joblin}, {Toublanc}, {Pech}, {Boissel},
  {Armengaud} \& {Frabel}}{{Joblin} et~al.}{prep}]{joblin09}
{Joblin} C.,  {Toublanc} D.,  {Pech} C.,  {Boissel} P.,  {Armengaud} M.,
  {Frabel} P.,  in prep.

\bibitem[\protect\citeauthoryear{{Jura}, {Balm} \& {Kahane}}{{Jura}
  et~al.}{1995}]{jura95}
{Jura} M.,  {Balm} S.~P.,    {Kahane} C.,  1995, \apj, 453, 721

\bibitem[\protect\citeauthoryear{{Kokkin}, {Troy}, {Nakajima}, {Nauta},
  {Varberg}, {Metha}, {Lucas} \& {Schmidt}}{{Kokkin} et~al.}{2008}]{kokkin08}
{Kokkin} D.~L.,  {Troy} T.~P.,  {Nakajima} M.,  {Nauta} K.,  {Varberg} T.~D.,
  {Metha} G.~F.,  {Lucas} N.~T.,    {Schmidt} T.~W.,  2008, \apjl, 681, L49

\bibitem[\protect\citeauthoryear{{Kroto}}{{Kroto}}{1988}]{kroto88}
{Kroto} H.,  1988, Science, 242, 1139

\bibitem[\protect\citeauthoryear{{Lafleur}, {Howard}, {Marr} \&
  {Yadav}}{{Lafleur} et~al.}{1993}]{lafleur93}
{Lafleur} A.~L.,  {Howard} J.~B.,  {Marr} J.~A.,    {Yadav} T.,  1993, J. Phys.
  Chem., 97, 13539

\bibitem[\protect\citeauthoryear{{Le Page}, {Keheyan}, {Snow} \&
  {Bierbaum}}{{Le Page} et~al.}{1999}]{lepage99}
{Le Page} V.,  {Keheyan} Y.,  {Snow} T.~P.,    {Bierbaum} V.~M.,  1999,
  International Journal of Mass Spectrometry and Ion Processes, 185, 949

\bibitem[\protect\citeauthoryear{{Le Page}, {Snow} \& {Bierbaum}}{{Le Page}
  et~al.}{2003}]{lepage03}
{Le Page} V.,  {Snow} T.~P.,    {Bierbaum} V.~M.,  2003, \apj, 584, 316

\bibitem[\protect\citeauthoryear{{L\'eger} \& {d'Hendecourt}}{{L\'eger} \&
  {d'Hendecourt}}{1985}]{leger85}
{L\'eger} A.,  {d'Hendecourt} L.,  1985, \aap, 146, 81

\bibitem[\protect\citeauthoryear{{L\'eger}, {d'Hendecourt} \&
  {Defourneau}}{{L\'eger} et~al.}{1989}]{leger89}
{L\'eger} A.,  {d'Hendecourt} L.,    {Defourneau} D.,  1989, \aap, 216, 148

\bibitem[\protect\citeauthoryear{{L\'eger} \& {Puget}}{{L\'eger} \&
  {Puget}}{1984}]{leger84}
{L\'eger} A.,  {Puget} J.~L.,  1984, \aap, 137, L5

\bibitem[\protect\citeauthoryear{{Li} \& {Draine}}{{Li} \&
  {Draine}}{2001}]{li01}
{Li} A.,  {Draine} B.~T.,  2001, \apj, 554, 778

\bibitem[\protect\citeauthoryear{{Lovas}, {McMahon}, {Grabow}, {Schnell},
  {Mack}, {Scott} \& {Luczkowsky}}{{Lovas} et~al.}{2005}]{lovas05}
{Lovas} F.~J.,  {McMahon} J.,  {Grabow} J.~U.,  {Schnell} M.,  {Mack} J.,
  {Scott} L.,    {Luczkowsky} L.,  2005, J. Am. Chem. Soc., 127, 4345

\bibitem[\protect\citeauthoryear{{Malloci}, {Joblin} \& {Mulas}}{{Malloci}
  et~al.}{2007}]{malloci07}
{Malloci} G.,  {Joblin} C.,    {Mulas} G.,  2007, Chemical Physics, 332, 353

\bibitem[\protect\citeauthoryear{{Men'shchikov}, {Schertl}, {Tuthill},
  {Weigelt} \& {Yungelson}}{{Men'shchikov} et~al.}{2002}]{menshchikov02}
{Men'shchikov} A.~B.,  {Schertl} D.,  {Tuthill} P.~G.,  {Weigelt} G.,
  {Yungelson} L.~R.,  2002, \aap, 393, 867

\bibitem[\protect\citeauthoryear{{Mulas}}{{Mulas}}{1998}]{mulas98}
{Mulas} G.,  1998, \aap, 338, 243

\bibitem[\protect\citeauthoryear{{Mulas}, {Malloci}, {Joblin} \&
  {Toublanc}}{{Mulas} et~al.}{2006a}]{mulas06c}
{Mulas} G.,  {Malloci} G.,  {Joblin} C.,    {Toublanc} D.,  2006a, \aap, 460,
  93

\bibitem[\protect\citeauthoryear{{Mulas}, {Malloci}, {Joblin} \&
  {Toublanc}}{{Mulas} et~al.}{2006b}]{mulas06b}
{Mulas} G.,  {Malloci} G.,  {Joblin} C.,    {Toublanc} D.,  2006b, \aap, 456,
  161

\bibitem[\protect\citeauthoryear{{Mulas}, {Malloci}, {Joblin} \&
  {Toublanc}}{{Mulas} et~al.}{2006c}]{mulas06a}
{Mulas} G.,  {Malloci} G.,  {Joblin} C.,    {Toublanc} D.,  2006c, \aap, 446,
  537

\bibitem[\protect\citeauthoryear{{Omont}}{{Omont}}{1986}]{omont86}
{Omont} A.,  1986, \aap, 164, 159

\bibitem[\protect\citeauthoryear{{Pech}, {Joblin} \& {Boissel}}{{Pech}
  et~al.}{2002}]{pech02}
{Pech} C.,  {Joblin} C.,    {Boissel} P.,  2002, \aap, 388, 639

\bibitem[\protect\citeauthoryear{{Peeters}, {Hony}, {Van Kerckhoven},
  {Tielens}, {Allamandola}, {Hudgins} \& {Bauschlicher}}{{Peeters}
  et~al.}{2002}]{peeters02}
{Peeters} E.,  {Hony} S.,  {Van Kerckhoven} C.,  {Tielens} A.~G.~G.~M.,
  {Allamandola} L.~J.,  {Hudgins} D.~M.,    {Bauschlicher} C.~W.,  2002, \aap,
  390, 1089

\bibitem[\protect\citeauthoryear{{Platt}}{{Platt}}{1956}]{platt56}
{Platt} J.~R.,  1956, \apj, 123, 486

\bibitem[\protect\citeauthoryear{{Rapacioli}, {Joblin} \&
  {Boissel}}{{Rapacioli} et~al.}{2005}]{rapacioli05}
{Rapacioli} M.,  {Joblin} C.,    {Boissel} P.,  2005, \aap, 429, 193

\bibitem[\protect\citeauthoryear{{Rouan}, {Leger}, {Omont} \& {Giard}}{{Rouan}
  et~al.}{1992}]{rouan92}
{Rouan} D.,  {Leger} A.,  {Omont} A.,    {Giard} M.,  1992, \aap, 253, 498

\bibitem[\protect\citeauthoryear{{Sarre}, {Miles} \& {Scarrott}}{{Sarre}
  et~al.}{1995}]{sarre95}
{Sarre} P.~J.,  {Miles} J.~R.,    {Scarrott} S.~M.,  1995, Science, 269, 674

\bibitem[\protect\citeauthoryear{{Scarrott}, {Watkin}, {Miles} \&
  {Sarre}}{{Scarrott} et~al.}{1992}]{scarrott92}
{Scarrott} S.~M.,  {Watkin} S.,  {Miles} J.~R.,    {Sarre} P.~J.,  1992,
  \mnras, 255, 11P

\bibitem[\protect\citeauthoryear{{Schmidt}, {Cohen} \& {Margon}}{{Schmidt}
  et~al.}{1980}]{schmidt80}
{Schmidt} G.~D.,  {Cohen} M.,    {Margon} B.,  1980, \apjl, 239, L133

\bibitem[\protect\citeauthoryear{{Surin}, {Dumesh}, {Lewen}, {Roth},
  {Kostromin}, {Rusin}, {Winnewisser} \& {Pak}}{{Surin} et~al.}{2001}]{surin01}
{Surin} L.~A.,  {Dumesh} B.~S.,  {Lewen} F.,  {Roth} D.~A.,  {Kostromin} V.~P.,
   {Rusin} F.~S.,  {Winnewisser} G.,    {Pak} I.,  2001, Review of Scientific
  Instruments, 72, 6

\bibitem[\protect\citeauthoryear{{Sygula}, {Xu}, {Marcinow} \&
  {Rabideau}}{{Sygula} et~al.}{2001}]{sygula01}
{Sygula} A.,  {Xu} G.,  {Marcinow} Z.,    {Rabideau} P.,  2001, Tetrahedron,
  57, 3637

\bibitem[\protect\citeauthoryear{{Thaddeus}}{{Thaddeus}}{2006}]{thaddeus06}
{Thaddeus} P.,  2006, Phil. Trans. R. Soc., 361, 1681

\bibitem[\protect\citeauthoryear{{Thorwirth}, {Theul{\'e}}, {Gottlieb},
  {McCarthy} \& {Thaddeus}}{{Thorwirth} et~al.}{2007}]{thorwirth07}
{Thorwirth} S.,  {Theul{\'e}} P.,  {Gottlieb} C.~A.,  {McCarthy} M.~C.,
  {Thaddeus} P.,  2007, \apj, 662, 1309

\bibitem[\protect\citeauthoryear{{Tielens}}{{Tielens}}{2005}]{tielens05}
{Tielens} A.~G.~G.~M.,  2005, {The Physics and Chemistry of the Interstellar
  Medium}.
The Physics and Chemistry of the Interstellar Medium, by A.~G.~G.~M.~Tielens,
  pp.~.~ISBN 0521826349.~Cambridge, UK: Cambridge University Press, 2005.

\bibitem[\protect\citeauthoryear{{van der Zwet} \& {Allamandola}}{{van der
  Zwet} \& {Allamandola}}{1985}]{vanderzwet85}
{van der Zwet} G.~P.,  {Allamandola} L.~J.,  1985, \aap, 146, 76

\bibitem[\protect\citeauthoryear{{Van Winckel}, {Cohen} \& {Gull}}{{Van
  Winckel} et~al.}{2002}]{vanwinckel02}
{Van Winckel} H.,  {Cohen} M.,    {Gull} T.~R.,  2002, \aap, 390, 147

\bibitem[\protect\citeauthoryear{{Vijh}, {Witt} \& {Gordon}}{{Vijh}
  et~al.}{2004}]{vijh04}
{Vijh} U.~P.,  {Witt} A.~N.,    {Gordon} K.~D.,  2004, \apjl, 606, L65

\bibitem[\protect\citeauthoryear{{Vijh}, {Witt} \& {Gordon}}{{Vijh}
  et~al.}{2005}]{vijh05}
{Vijh} U.~P.,  {Witt} A.~N.,    {Gordon} K.~D.,  2005, \apj, 633, 262

\bibitem[\protect\citeauthoryear{{Waters}, {Waelkens}, {van Winckel},
  {Molster}, {Tielens}, {van Loon}, {Morris}, {Cami}, {Bouwman}, {de Koter},
  {de Jong} \& {de Graauw}}{{Waters} et~al.}{1998}]{waters98}
{Waters} L.~B.~F.~M.,  {Waelkens} C.,  {van Winckel} H.,  {Molster} F.~J.,
  {Tielens} A.~G.~G.~M.,  {van Loon} J.,  {Morris} P.~W.,  {Cami} J.,
  {Bouwman} J.,  {de Koter} A.,  {de Jong} T.,    {de Graauw} T.,  1998,
  Nature, 391, 868

\end{thebibliography}

\appendix
\clearpage
\section{Online material}
\begin{figure}
\begin{center}
\includegraphics[trim=2cm 0cm 0.5cm 2.5cm , clip,  angle=-90, width=0.45 \hsize]{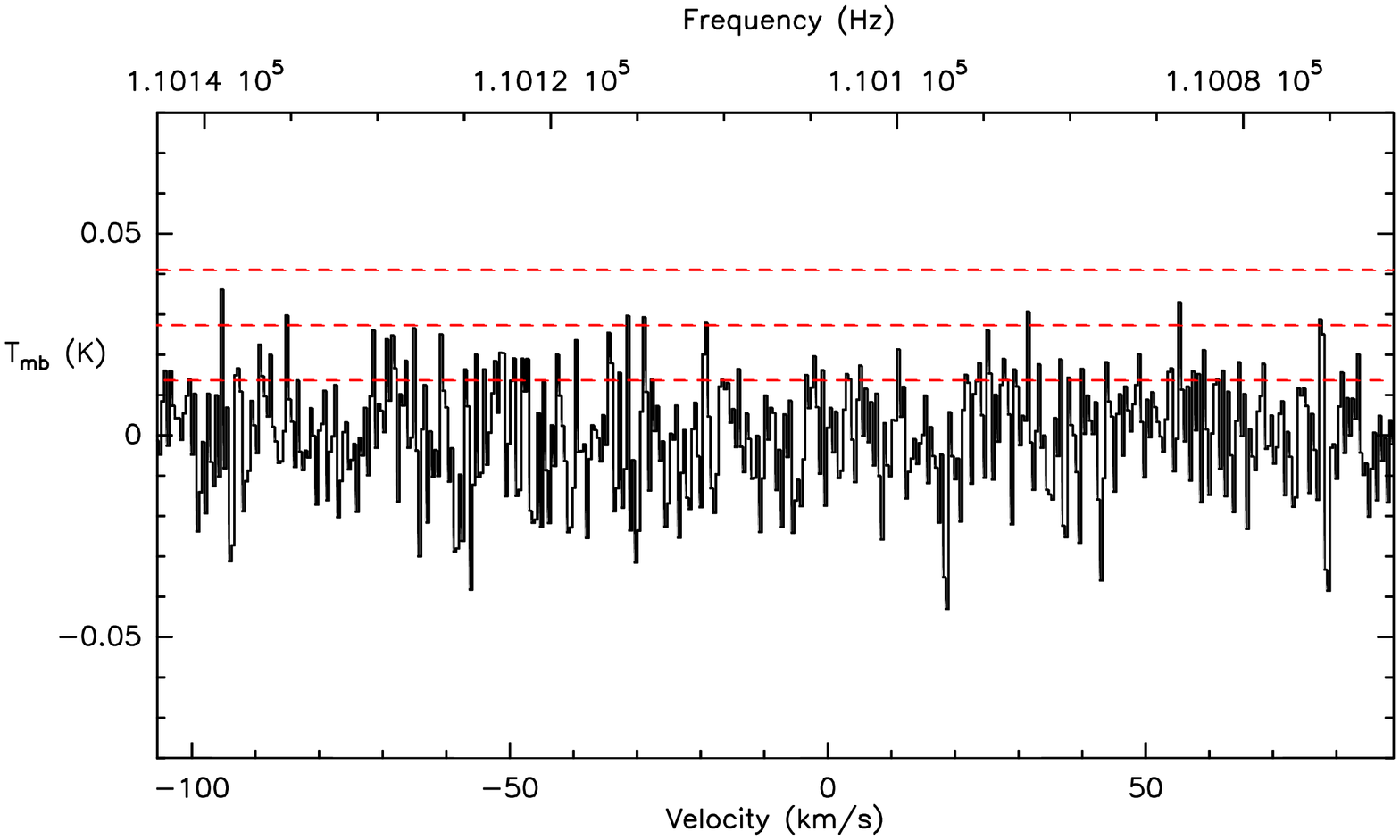} \hfill
\includegraphics[trim=2cm 0cm 0.5cm 2.5cm , clip,  angle=-90, width=0.45\hsize]{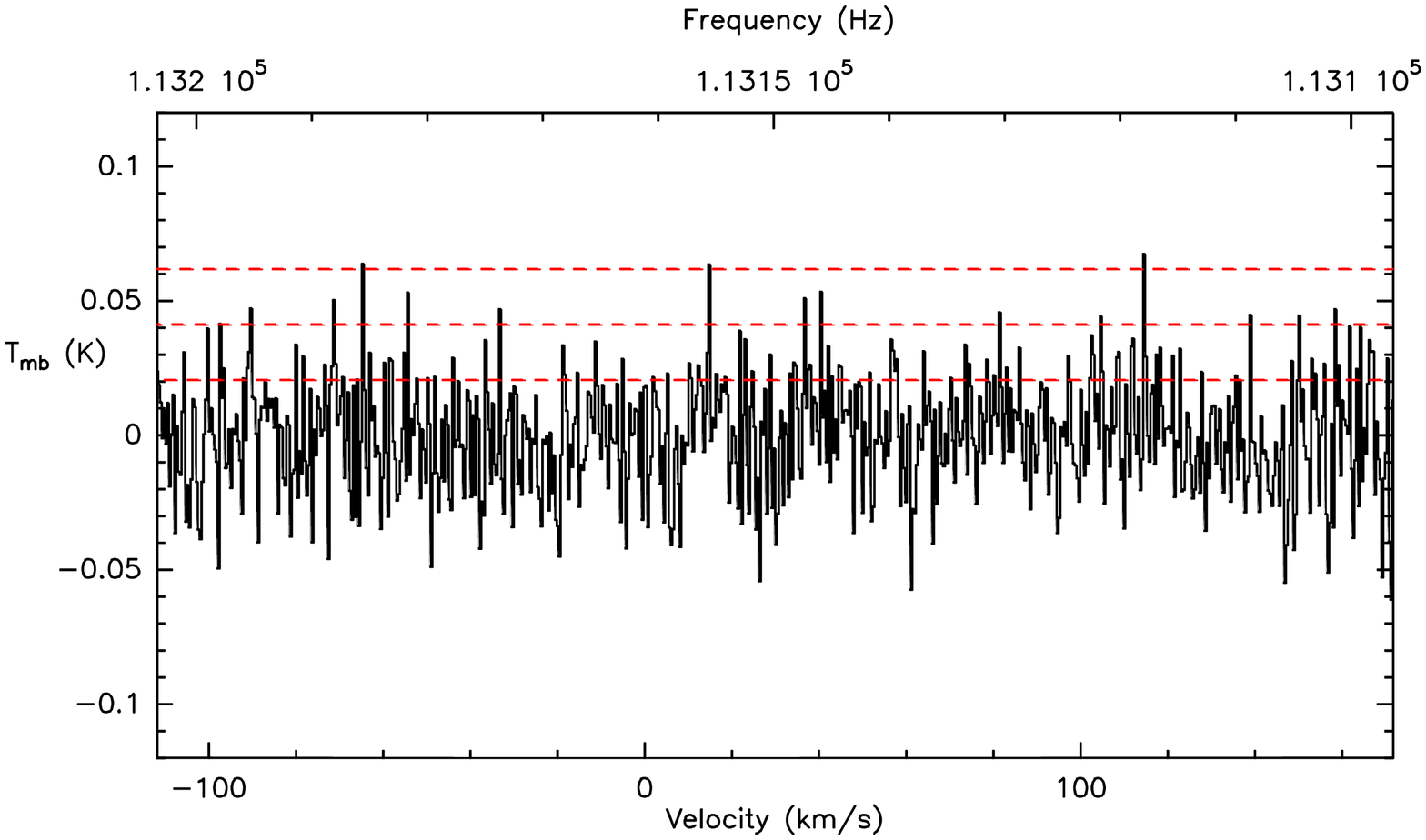}
\includegraphics[trim=2cm 0cm 0.5cm 2.5cm , clip,  angle=-90, width=0.45\hsize]{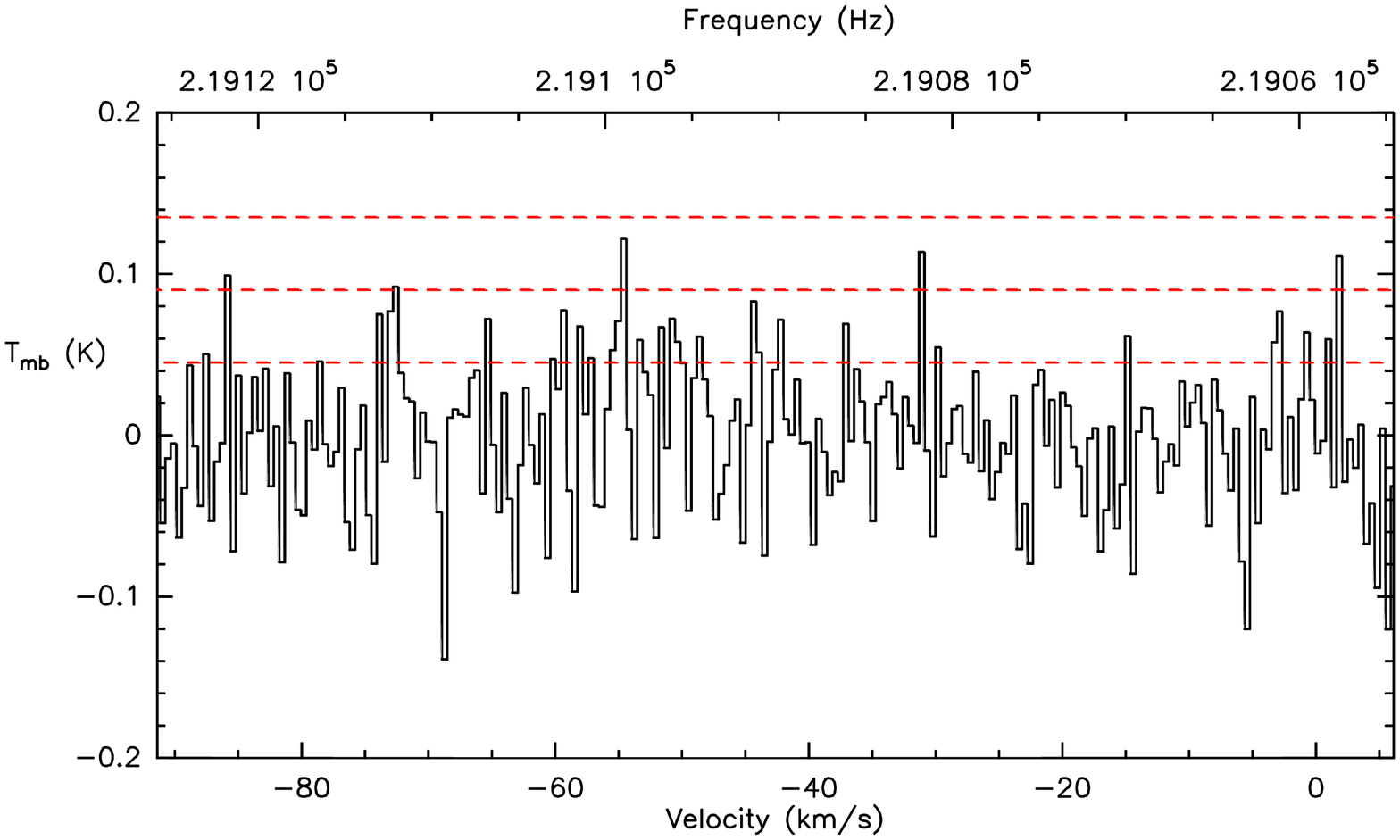} \hfill
\includegraphics[trim=2cm 0cm 0.5cm 2.5cm , clip,  angle=-90, width=0.45\hsize]{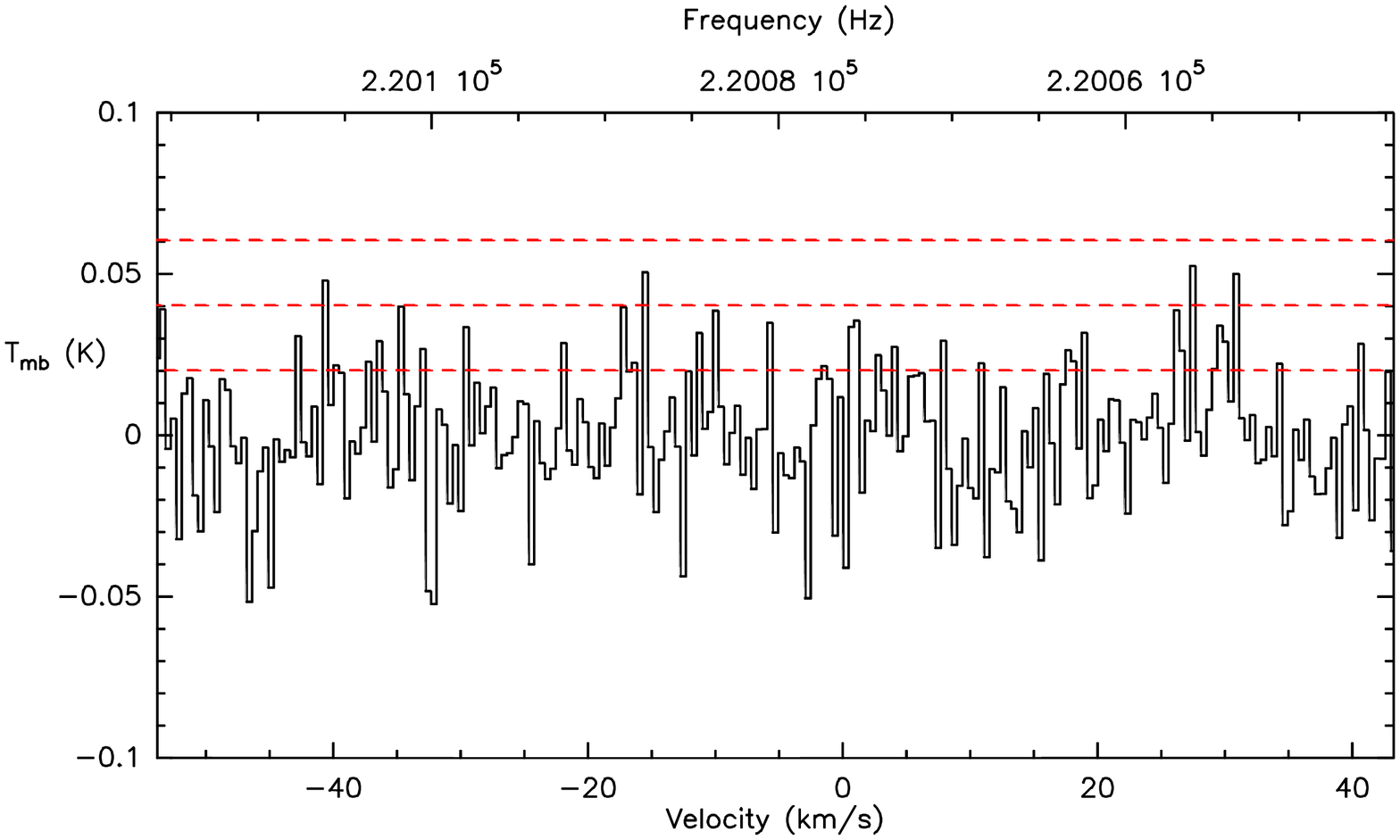}
\end{center}
\caption{Observations of the expected frequency range for the pure rotational transitions of corannulene J +1 $\rightarrow$ J: 108 $\rightarrow$ 107, 111 $\rightarrow$ 110 and 215 $\rightarrow$ 214 and 216 $\rightarrow$ 215. }
\end{figure}
\begin{figure}
\begin{center}
\includegraphics[trim=2cm 0cm 0.5cm 2.5cm , clip,  angle=-90, width=0.45\hsize]{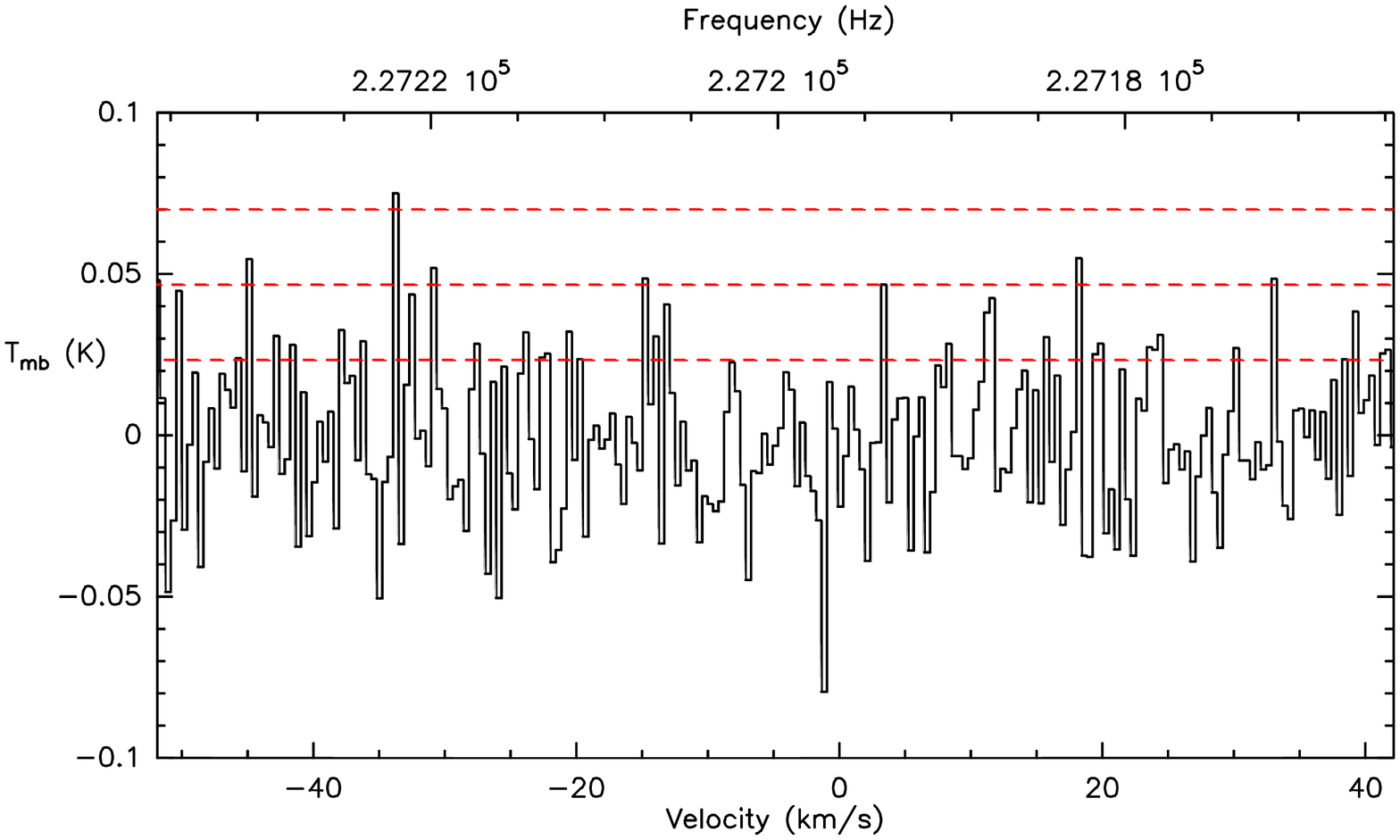} \hfill
\includegraphics[trim=2cm 0cm 0.5cm 2.5cm , clip,  angle=-90, width=0.45\hsize]{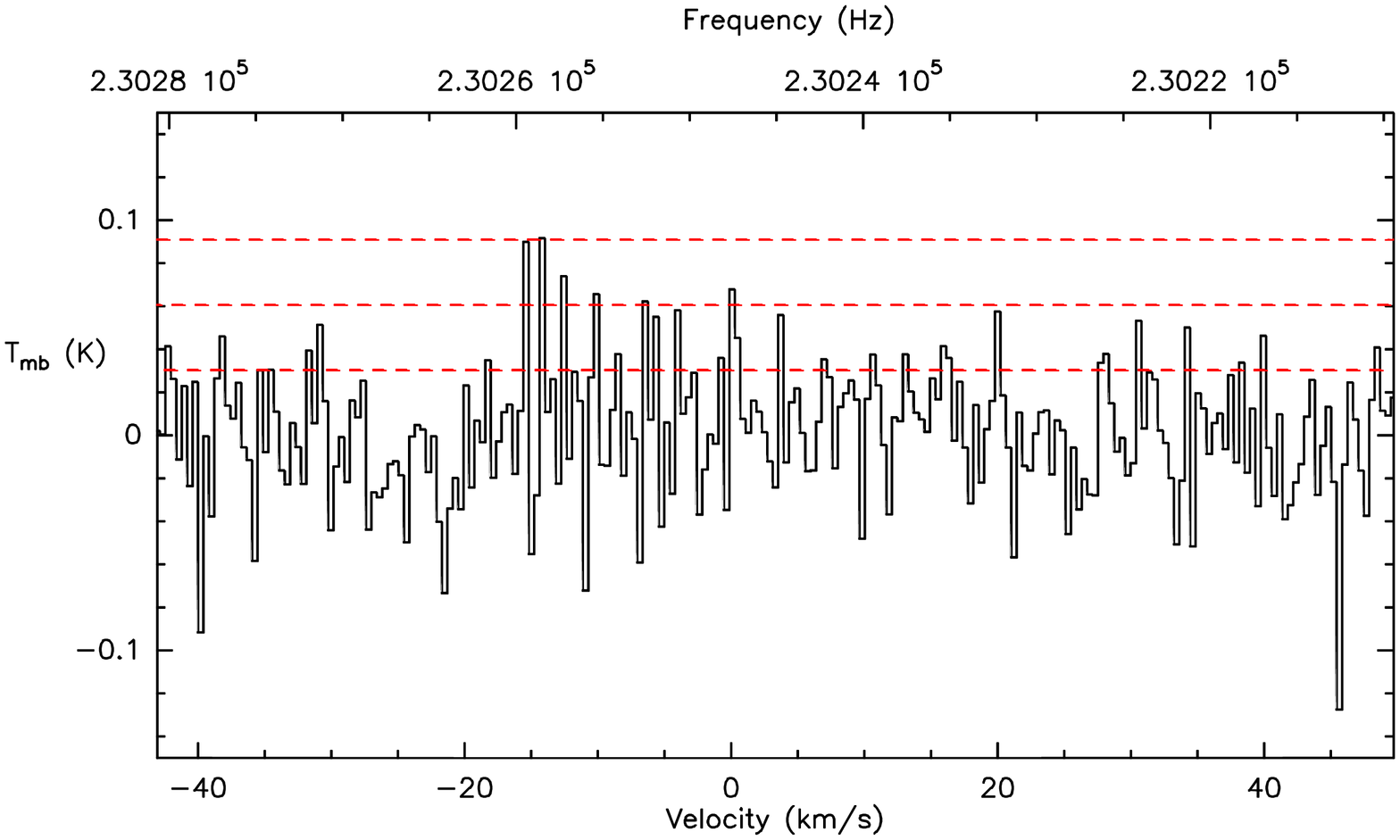} 
\includegraphics[trim=2cm 0cm 0.5cm 2.5cm , clip,  angle=-90, width=0.45\hsize]{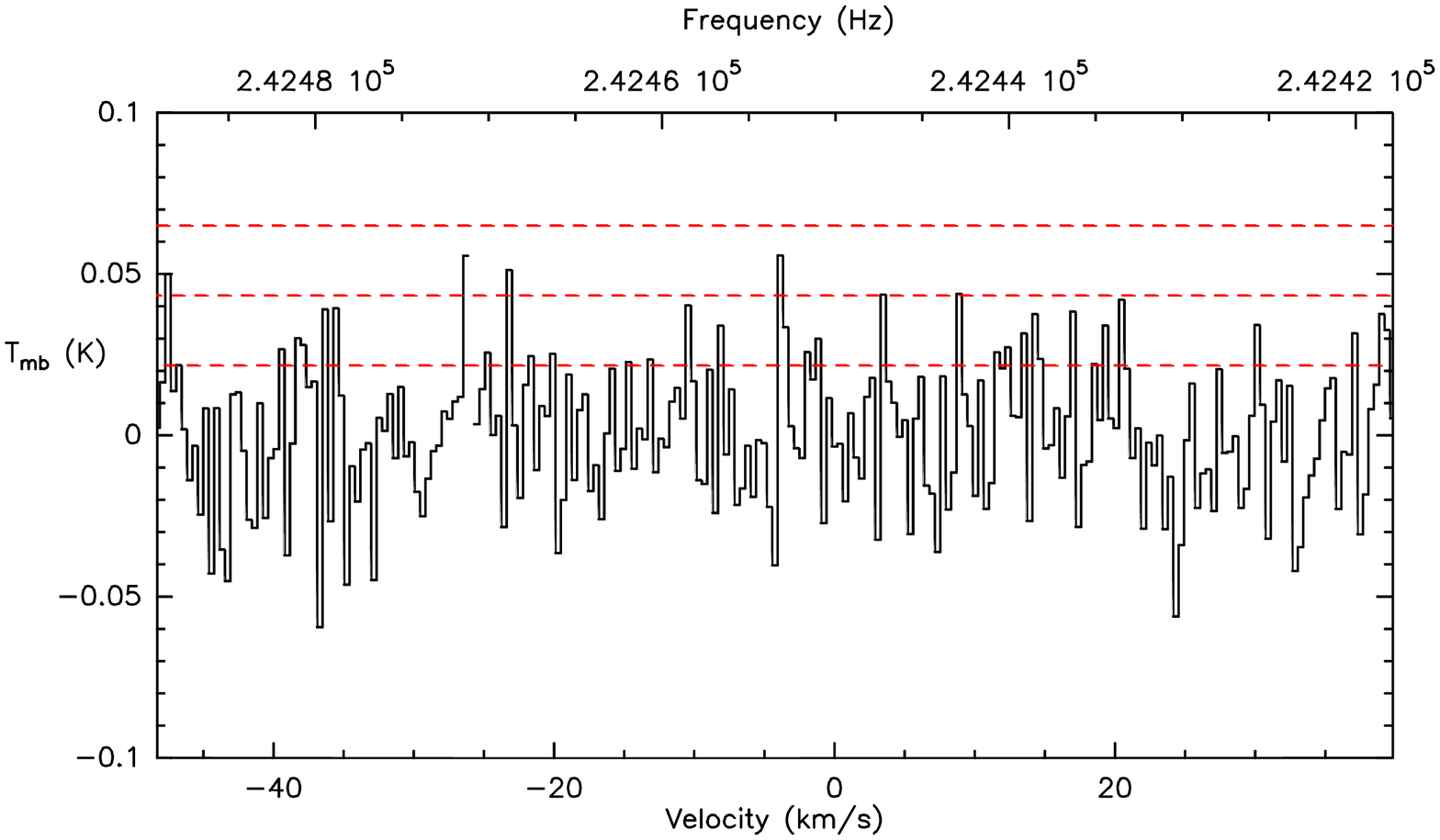} \hfill
\includegraphics[trim=2cm 0cm 0.5cm 2.5cm , clip,  angle=-90, width=0.45\hsize]{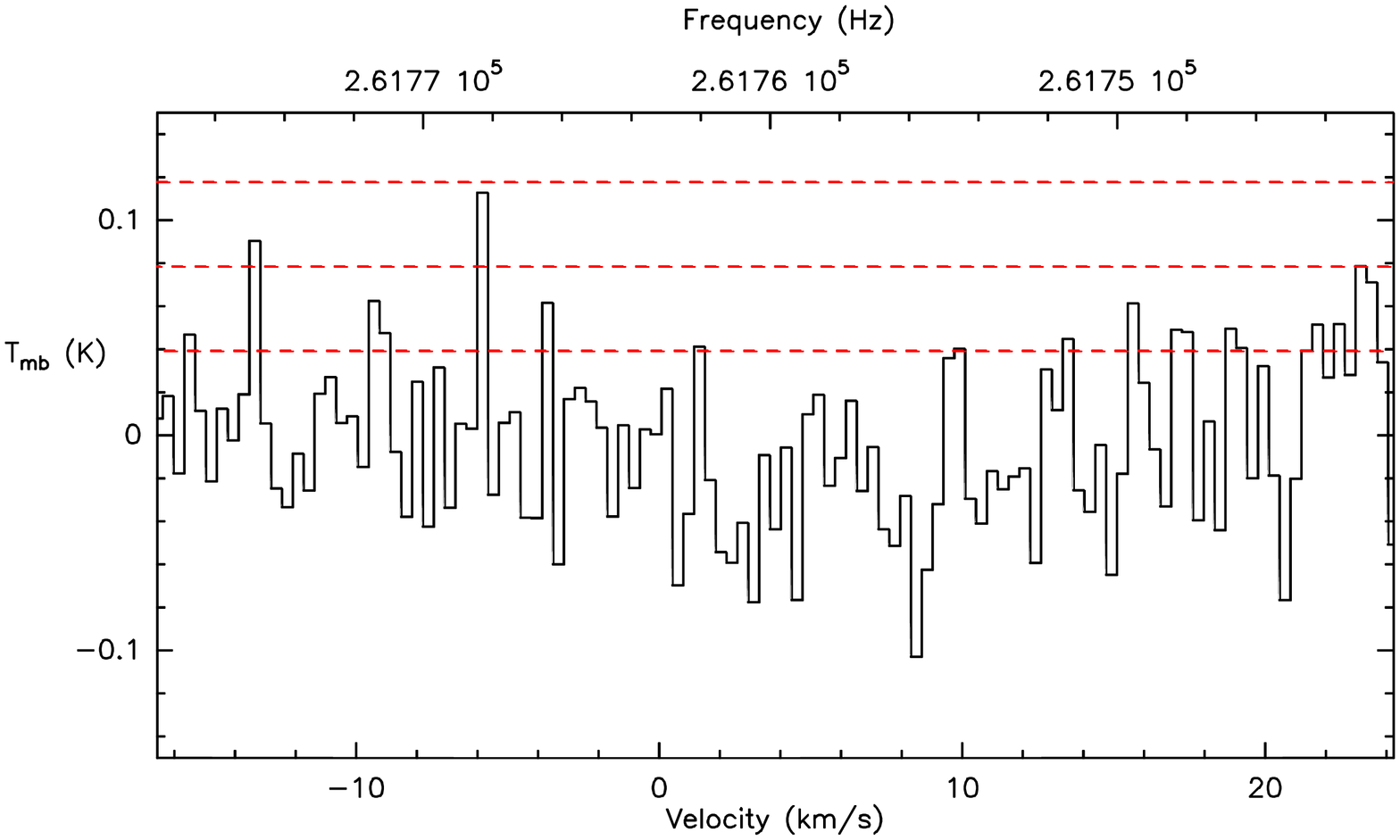}
\end{center}
\caption{Observations of the expected frequency range for the pure rotational transitions of corannulene J +1 $\rightarrow$ J:  223 $\rightarrow$ 222,  226 $\rightarrow$ 225, 238 $\rightarrow$ 237 and 257 $\rightarrow$ 256. }
\end{figure}

\end{document}